\documentclass[a4paper, 10pt]{amsart}

\pdfoutput=1
\numberwithin{equation}{section} \oddsidemargin=-.0cm
\usepackage[utf8]{inputenc}
\usepackage{amssymb, amsmath}
\usepackage{amsfonts}
\usepackage{float}
\usepackage{graphics}
\usepackage{subfigure}

\usepackage{color}
\usepackage{graphicx}
\usepackage{enumitem}
\usepackage[unicode, colorlinks, pdfpagelabels]{hyperref}
\usepackage{tikz}
\usepackage{pgfplots}
\pgfplotsset{compat=newest}
\usetikzlibrary{arrows}
\usetikzlibrary{decorations.text}
\usetikzlibrary{decorations.markings}
\usepackage{pgfplotstable}
\usepackage{appendix}

\usepackage{filecontents}

\newcommand{\abs}[1]{\lvert #1 \rvert}

\renewcommand{\Re}{\operatorname{Re}}
\renewcommand{\Im}{\operatorname{Im}}

\definecolor{greenrb}{rgb}{0.2,0.6,0.2}
\definecolor{rred}{rgb}{0.7,0,0.1}

\def\d{\, \mathrm{d}}

\newtheorem{theorem}{Theorem}[section]
\newtheorem{lemma}{Lemma}[section]
\newtheorem{remark}{Remark}[section]

\title[Two Layer QG Model]{Transitions of zonal flows in a two-layer quasi-geostrophic ocean model}
\author[Chekroun]{Mickael D.~Chekroun}
\address[Chekroun]{Department of Earth and Planetary Sciences, Weizmann Institute,
Rehovot 76100, Israel}
\email{michael-david.chekroun@weizmann.ac.il}

\author[Dijkstra]{Henk Dijkstra}
\address[Dijkstra]{Centre for Complex System Studies, Department of Physics,
Utrecht University, Utrecht, The Netherlands}
\email{h.a.dijkstra@uu.nl}

\author[Şengül]{Taylan Şengül}
\address[Şengül]{Department of Mathematics, Marmara University, 34722 Istanbul, Turkey}
\email{taylan.sengul@marmara.edu.tr}

\author[Wang]{Shouhong Wang}
\address[Wang]{Department of Mathematics,
Indiana University, Bloomington, IN 47405}
\email{showang@indiana.edu, http://www.indiana.edu/\~{ }fluid}

\date{\today}

\begin{document}

\begin{abstract}
  We consider a 2-layer quasi-geostrophic ocean model where the upper layer
  is forced by a steady Kolmogorov wind stress in a periodic channel domain,
  which allows to mathematically study the nonlinear development of the
  resulting flow.  The model supports a steady parallel shear flow as a response to the
  wind stress. As the maximal velocity of the shear flow (equivalently the maximal
  amplitude  of the wind forcing) exceeds a critical threshold, the zonal jet destabilizes
  due to baroclinic instability and we numerically demonstrate  that a first transition
  occurs. We obtain reduced equations of the system using the formalism of dynamic
  transition theory and establish two scenarios  which completely describe this first
  transition. The generic scenario is that {a conjugate pair of modes loses stability} and a Hopf
  bifurcation occurs as a result. Under an appropriate set of parameters describing
  related midlatitude oceanic flows, we show that this first transition is continuous: a supercritical
  Hopf bifurcation occurs and a stable time periodic solution bifurcates.  We also investigate
  the case of double Hopf bifurcations which occur when four modes of the linear stability
  problem  simultaneously destabilize the zonal jet.  In this case we prove that,  in the relevant
  parameter regime, the flow exhibits a continuous transition accompanied by a bifurcated
  attractor homeomorphic to $S^3$. The topological structure of this attractor is analyzed in
  detail and is shown to depend on the system parameters. In particular, this attractor contains
   (stable or unstable)  time-periodic solutions and a quasi-periodic solution.
\end{abstract}

\maketitle
\tableofcontents

\section{Introduction}~\label{sec:Introduction}

Baroclinic instability is among the most important geophysical fluid dynamical instabilities playing a crucial
role in the dynamics of atmospheres and oceans. In particular, this  instability mechanism is the dominant process in atmospheric dynamics
shaping the cyclones and anticyclones that dominate weather in mid-latitudes, as well as the mesoscale ocean eddies
that play various roles in oceanic dynamics and the transport of heat and salt \cite{Vallis2006}.
Much is known on the linear stability of zonal jets in a horizontally unbounded ocean in the quasi-geostrophic
(QG) flow regime. Classical models, such as the continuously stratified Eady model \cite{Eady1949} and the
two-layer Phillips model \cite{Phillips1951}, have lead to a detailed  understanding of the mechanism of
baroclinic instability of a zonal   jet in the inviscid case.  Long waves destabilize the zonal jet with maximum
growth rates occurring for  perturbations  having wavelengths on the order of the Rossby deformation radius,
typically $50-100$ km for the mid-latitude ocean \cite{Talley2011}.

In  case linear friction is included in the two-layer model, the neutral curve has a minimum at $(k_c, \mu_c)$
where $k_c$ is the  critical wavenumber and $\mu_c$ the critical value of the control parameter (e.g. the
maximum speed of the zonal jet).  The nonlinear development of these perturbations has been extensively
analyzed in the weakly nonlinear case \cite{pedlosky1970,Romea1977,welander1966,lovegrove2002}.
In the regime $| k - k_c | = {\mathcal O}(\epsilon)$ and $| \mu - \mu_c | = {\mathcal O}(\epsilon^2)$,
\cite{VanderVaart1997} showed that on a long time scale $T = \epsilon^2 t$ and large spatial scale
$X = \epsilon(x - c_g t)$, where $c_g$ is the group velocity of the waves at criticality, the complex
amplitude $A$ of the wave packet destabilizing the jet satisfies a Ginzburg-Landau equation, written as
\begin{displaymath}
\frac{\partial A}{\partial T} = \gamma_1 A + \gamma_2 \frac{\partial^2 A}{\partial X^2}
- \gamma_3 A |A|^2,
\label{e:GLE}
\end{displaymath}
where the $\gamma_i$ are complex constants.
\cite{VanderVaart1997} also showed that the fixed point solution of this equation can
become unstable to sideband instabilities. Subsequent analysis has shown \cite{Dijkstra1998}
that upgradient momentum transport can occur due to the self-interaction of the instabilities
leading to rectification of the zonal jet.

In reality, the ocean basins are zonally bounded by continents and the midlatitude zonal jets are
part of the gyre system, for example the subpolar gyre and subtropical gyre in the North Atlantic,
forced by the surface wind stress through Ekman pumping \cite{pedlosky1987}.  The problem of
baroclinic instability  of such non-parallel flows is much more complicated and has so far only been
tackled numerically.  When the wind-forced QG equations are  discretized,  the linear stability problem
for the gyre flow  results in a  large-dimensional generalized  eigenvalue problem, typically of
dimension $10^4$.  There are many results for the one-layer single- and double-gyre flows (for
an overview, see \cite[Chap.~5]{Dijkstra2005}), but in  this case there is no baroclinic instability.
There are relatively few  results for the  two-layer case. In  \cite{Dijkstra1997}, it was shown that in
the two-layer case  the double-gyre flow becomes unstable through  a sequence of Hopf bifurcations.
The  perturbation flow patterns at criticality are ``banana-shaped,''  locally resembling those of  baroclinic
instability in the Philips model. Stable periodic orbits result  from these Hopf bifurcations, typically
given rise to meandering motion of the gyre boundary.

As an intermediate, more analytically tractable case, we consider here the baroclinic instability of a
zonal jet for a two-layer QG model in a zonally periodic channel. In this case, the properties of the
bounded geometry are somehow represented, as the patterns of the unstable modes are restricted
by the periodicity of the channel, so a sequence of Hopf bifurcations is expected just as in the more
realistic gyre case. In addition, parallel flow solutions exist in the zonally periodic channel which
simplifies the linear stability problem substantially such that a more detailed nonlinear analysis,
akin to that in the horizontally unbounded case, can be performed. The parallel flow can also be
connected to the surface wind stress, as in the full gyre case, but at the expense of adding
an additional linear friction term to the upper layer vorticity equation; for more details, see \autoref{Sec2} below. {In this way, the situation studied is
more relevant for the stability of the Antarctic Circumpolar Current, than for the midlatitude 
gyre circulations.} 

The case specifically studied in the paper is the circulation set up by a time-independent Kolmogorov
wind-stress field (for $k = 1, 2, \ldots$)
\begin{displaymath}
  \tau^{x}(y) = -  \tau \frac{\tau_0}{k \pi}  \cos k \pi \frac{y}{L_y} ~ ; ~  \tau^{y} = 0
\end{displaymath}
where $\tau_0$ is a characteristic mid-latitude wind-stress value. This wind stress
forces an ocean in a periodic channel $[0,\mathbb R/(2L_x \mathbb Z)] \times [-L_y,L_y]$
on the $\beta$-plane. The case $k=1$ and $k=2$ are often referred to as the single- and
double-gyre forcing. The stratification is modeled in terms of a two-layer system and the wind
stress only directly forces the upper layer. As a response to this wind stress, the system supports
a basic shear flow $\psi^s$. The amplitude $\tau$ that controls the wind-stress curl, or
equivalently the maximal velocity of the shear flow $\psi^s$ is chosen as the bifurcation
parameter.

We first perform a numerical linear stability analysis of this basic shear flow;  for  small values of
$\tau$, all associated  eigenvalues have negative real parts  such that the jet is stable.   When the
aspect ratio of the channel $a = L_y/L_x$ is large, the eigenvalues remain in the left complex plane
regardless of the value of $\tau$. However, when the aspect ratio gets small, the basic shear flow loses
stability at a critical $\tau$ in the form of a {single conjugate pair or two conjugate pairs of} eigenvalues crossing the
imaginary axis, giving rise to {either} a Hopf or a double Hopf bifurcation. We next  use the idea and method  of the dynamic transition theory \cite{ma2005,ptd}, which is aimed to determine all the local attractors near
a transition.  The approach  {allows for} a classification {near the instability onset} of all transitions into three classes known as
continuous,  catastrophic and random types \cite{ptd}. In this way, our study extends previous results using this approach
on the single-layer barotropic case~\cite{simonnet2003,dijkstra2015}, the two-layer case for
constant zonal jet velocities ~\cite{cai2017} and the barotropic Munk western boundary layer
current profile case \cite{kieu2018}, to the zonally periodically bounded two-layer case.

Using the center manifold reduction, we obtain {effective} reduced (ordinary differential equation, ODE) models
describing this transition. {The dynamic transition theory identifies then transition numbers that qualifies the transition's type and are calculated from the reduced ODE's coefficients}.
The case of a Hopf bifurcation is generic while the case of a
double Hopf bifurcation is degenerate and requires fine tuning of the aspect ratio to critical values
where two {conjugate} pairs of eigenmodes with consecutive wavenumbers {have their eigenvalues crossing} the imaginary axis
simultaneously.  Using standard parameter values describing the {midlatitude related oceanic flows}, we perform
numerical computations of the transition number for the forcing patterns corresponding to $k=1, 2, 3$
and the aspect ratios $a \ge 3$. We find that in the parameter regimes we are interested in, the Hopf
bifurcation is supercritical and a stable limit cycle bifurcates. For the double Hopf bifurcation, we
find that after the corresponding transition takes place, the system exhibits a bifurcated local attractor ~\cite{ma2005} near
the basic shear flow which is homeomorphic to the 3D-sphere. The topological structure of this
attractor is analyzed and depending on the parameters, it is found to contain a combination of
limit cycles and a quasi periodic solution.

The paper is organized as follows. In \autoref{Sec2}, the quasi-geostrophic model is presented.
This is followed by \autoref{Sec_results} where the theory and numerical results for the linear stability
problem (\autoref{sec:LIA}), the Hopf bifurcation case (\autoref{Sec_Hopf}) and the double-Hopf bifurcation
case (\autoref{Sec_doubleHopf}) are presented. These results are summarized and discussed in \autoref{Sec_discuss}.
{\autoref{Sec_AppB} and \autoref{Sec_AppD} contains details regarding the proofs of the transition theorems and the explicit formulas of the corresponding reduced ODE's coefficients, in the Hopf and double-Hopf cases, respectively. \autoref{Sec_AppA} and \autoref{Sec_AppC}, as for them, provide details about  for the numerical treatment of the linear stability analysis and the practical computation of the transition numbers. Finally the set of model parameters used in the numerical study of the problem is given in \autoref{Sec_AppE}. }

\section{The model}\label{Sec2}
We consider two layers of homogeneous fluids, each with a different and constant density
$\rho_1$ and $\rho_2$ and with equilibrium layer thicknesses $H_1$ and $H_2$, on a
mid-latitude $\beta$-plane with Coriolis parameter $f = f_{0} + \beta_0 y$.
The lighter fluid in layer 1 is assumed to lie on top of the heavier one in layer 2 so that
the stratification is  statically stable, i.e., $\rho_1 < \rho_2$; bottom topography is neglected.

This flow can be modeled by the two-layer QG model \cite{pedlosky1970} using the
geostrophic stream function $\psi_i$ and the vertical component of the relative
vorticity $\zeta_i$ in each layer ($i = 1,2$). The quantities $\psi_i$ and $\zeta_i$
are non-dimensionalised by $U L_y$ and $U/L_y$, respectively, wind stress with
$\tau_0$, length with $L_y$, and time with $L_y/U$, where $U$ is a characteristic
horizontal velocity. { By choosing} $U = \tau_0/(\rho_0 \beta_0 L_y H_1)$, where $\rho_0$
is a reference density,  the dimensionless equations on the domain
$(0, 2/a) \times (-1, 1)$ {(with $a = L_y/L_x$)} become
\begin{equation}
  \label{eq:main-1}
  \begin{aligned}
    & \left[ \frac{\partial }{\partial t} + \{ \psi_{1}, \cdot \} \right]
    \left( \Delta \psi_{1} + F_{1} ( \psi_{2} - \psi_{1} ) + \beta y \right) =
    -r_1 \Delta \psi_1  - \tau \beta \sin k \pi y\\
    & \left[ \frac{\partial }{\partial t} + \{ \psi_{2}, \cdot \} \right]
    \left( \Delta \psi_{2} + F_{2} ( \psi_{1} - \psi_{2} ) + \beta y \right) =
    -r_{2} \Delta \psi_{2},
  \end{aligned}
\end{equation}
where $\{ f, g\} = f_{x} g_{y} - f_{y} g_{x}$ is the usual Jacobian operator and
\[
{\mathcal F}_1 = - r_1 \Delta \psi_{1},
\]
represents the damping  of upper layer vorticity due to frictional
processes.  In the bottom layer, we include a linear (Ekman)
friction term  $- r_{2} \Delta \psi_2$; in both layers, Laplacian friction terms are
neglected { due to the absence of} continental boundary layers { making such terms
much smaller than the other ones}. The expressions for the dimensional and
dimensionless parameters, with their standard values at a latitude
$45^\circ$N,  are given in \autoref{tab:defp}.

For the boundary conditions, we assume periodicity in the $x$-direction
and free-slip boundaries in the $y$-direction. Hence, the conditions are
\begin{equation}
  \label{BC}
  \begin{aligned}
    & \psi_{i} \mid_{x=0} = \psi_{i} \mid_{x=2/a}, \qquad i=1,2. \\
    & \psi_{i} \mid_{y = \pm 1} = \frac{\partial^{2} \psi_{i}}{\partial y^{2}}
    \mid_{y = \pm 1} = 0, \qquad i=1,2.
  \end{aligned}
\end{equation}

In actual ocean basins, a steady zonal jet is generated by the applied wind stress
through Ekman pumping, a Sverdrup balance and a western boundary layer flow
\cite{pedlosky1987}. Due to the periodic boundary conditions used here, such a flow
cannot be captured in this model. However, {due to the presence of the large
upper layer friction, which corrects for the absence of the Sverdrup balance, the 
equations  allow} a steady state
of the form
\begin{equation}
  \label{eq:steady-state}
  \psi_{1}^{s} = \Psi \sin k \pi y, \qquad \psi_{2}^{s} = 0,
\end{equation}
which { relates} to the wind stress field, when  ${\mathcal F}_1 - \tau \beta \sin k \pi y = 0$.
In this paper, we will assume that the wind-stress vorticity  input is balanced by vorticity decay
due to the linear friction term ${\mathcal F}_1 = - r_1 \Delta \psi_{1}$, being aware that
a larger friction coefficient $r_1$ is needed than can be justified from existing dissipative
processes in the ocean.  In this case, it follows that
\begin{equation}
  \label{eq:1}
  \Psi = \frac{ \tau \beta}{ (k \pi)^2 r_1  }.
\end{equation}
The parameter $\Psi$ appearing in~\eqref{eq:steady-state} can then be
chosen as the control parameter as is the case in this study, instead of $\tau$.
{It can be interpreted as the maximal (zonal) velocity of the shear flow (after taking the derivative in $y$ of \eqref{eq:steady-state}) or equivalently the maximal
  amplitude  of the wind forcing according to \eqref{eq:1}.}

By considering the perturbation $\psi_{i}' = \psi_{i} - \psi^{s}_{i}$, $i=1,2$, we can write the system
{into} the following operator form {(after dropping the prime notation)}
\begin{equation}
  \label{eq:main-abstract-form}
  \mathcal{M} \partial_{t} \psi = \mathcal{N} \psi + \mathcal{G}(\psi), \qquad \psi = (\psi_{1}, \psi_{2}),
\end{equation}
where $\mathcal{M}$ and $\mathcal{N}$ are the linear operators defined as
\begin{equation}
  \label{eq:M}
  \mathcal{M} \psi =
  \begin{bmatrix}
    \Delta \psi_{1} + F_{1}( \psi_{2} - \psi_{1} ) \\
    \Delta \psi_{2} + F_{2}( \psi_{1} - \psi_{2} )
  \end{bmatrix},
\end{equation}
{and}
\begin{equation}
  \label{eq:N}
  \mathcal{N} \psi =
  \begin{bmatrix}
    \Psi k \pi \cos k \pi y \left( (k \pi)^{2} \frac{\partial \psi_{1}}{\partial x} + F_{1} \frac{\partial \psi_{2}}{\partial x} + \frac{\partial \Delta \psi_{1}}{\partial x}  \right) - \beta \frac{\partial \psi_{1}}{\partial x} - r_{1} \Delta \psi_{1} \\
    -\Psi k \pi  F_{2} {\cos k \pi y} \frac{\partial \psi_{2}}{\partial x} - \beta \frac{\partial \psi_{2}}{\partial x} - r_{2} \Delta \psi_{2} \\
  \end{bmatrix}.
\end{equation}
Lastly, the bilinear nonlinearity is given { explicitly by}
\begin{equation}
  \label{G}
  \mathcal{G}(\psi) =
  \begin{bmatrix}
    & - \{ \psi_{1}, \Delta \psi_{1} + F_{1} (\psi_{2} - \psi_{1}) \} \\
    & - \{ \psi_{2}, \Delta \psi_{2} + F_{2} (\psi_{1} - \psi_{2}) \}
  \end{bmatrix}.
\end{equation}

{The operators $\mathcal{G}$ and $\mathcal{N}$ are well-defined mappings, $\mathcal{G}: H_1\to H_{-1}$ and $\mathcal{N}:H_0\to H_{-1}$, on the following functional spaces:}
\begin{equation} \label{Func_spaces}
\begin{aligned}
& H_{1} = \{\psi = (\psi_1, \psi_2) \in H^{4}(\Omega)\times H^{4}(\Omega)
|~\psi~\text{satisfies}~\eqref{BC}\},\\
& H_{0} = \{\psi = (\psi_1, \psi_2) \in H^{2}(\Omega)\times H^{2}(\Omega)
|~\psi~\text{satisfies}~\eqref{BC}\}, \\
& H_{-1} = L^2(\Omega)^2.
\end{aligned} 
\end{equation}
Here, 
\begin{equation}\label{Eq_Omega}
{\Omega = (0,2/a)\times (-1, 1), \textrm{ with } a=L_y/L_x,}
\end{equation}
and $H^{4}(\Omega)$, $H^{2}(\Omega)$, $L^2(\Omega)$ are the usual Sobolev and Lebesgue function spaces endowed with their natural inner products. These functional spaces account for spatial regularity of the solution $\psi$ for which $H^{p}(\Omega)$ denotes the space of square-integrable functions that possess $p$th-order derivatives (in the distribution sense) that are themselves square-integrable; see e.g.~\cite{brezis_book}.
{Using this functional framework, it can be shown that Eq.~\eqref{eq:main-abstract-form}  is well-posed by application of e.g.~a standard approximation method (such as Galerkin) \cite{bernier1992existence}.}


\section{Results}\label{Sec_results}

In this section, we first present the linear stability analysis of the basic shear flow and then we move on to describe the first transitions due to the instabilities, covering both the Hopf and double Hopf bifurcations. {Although in realistic ocean basins the aspect ratio $a$ would be small, we allow here the full range of $a$ to also study the nonlinear interactions of localized instabilities.}
\vspace{2mm}

\subsection{Linear Stability Analysis}~\label{sec:LIA}
We first investigate the linear stability of the basic solution.
For this purpose, we denote the eigenmodes of the linear problem by
\begin{displaymath}
  \psi_{m, j}(x, y) = e^{i \alpha_{m} x} Y_{j}(y), \qquad j \in \mathbb{N}, m \in \mathbb{Z}, \quad
  \alpha_{m} = a m \pi.
\end{displaymath}
with eigenvalues $\sigma_{m, j}$, i.e.
\begin{equation}
  \label{evp}
  \sigma_{m, j}\mathcal{M} \psi_{m, j} = \mathcal{N} \psi_{m, j}.
\end{equation}
Since the linear operators $\mathcal{M}$ and $\mathcal{N}$ are real, we have
\begin{displaymath}
  \overline{ \sigma_{m, j} } = \sigma_{-m, j}, \qquad
  \overline{ \psi_{m, j} } = \psi_{-m, j}, \qquad \forall m \in \mathbb{N}.
\end{displaymath}
\vspace{2mm}%
%

\begin{figure}[ht]%
  \centering
  \begin{tikzpicture}[]
    \begin{axis}[x=80mm, y=.6mm, ymin=-40, ymax=40, xmin=-1.2, xtick={-1, 0},
      xlabel=$\Re(\sigma_{m, j})$, ylabel=$\Im(\sigma_{m, j})$, legend
      style={at={(1.05,1.05)},anchor=north,legend cell align=left}]
      \addplot[mark size=1.5, only marks, mark=triangle, red] table [col sep=comma, x=rm2, y=im2] {eigenvalues.csv};
      \addlegendentry{\tiny $m=-2m_{c}$}
      \addplot[mark size=1.5, only marks, mark=+, blue] table [col sep=comma, x=rm1, y=im1] {eigenvalues.csv};
      \addlegendentry{\tiny $m=-m_{c}$}
      \addplot[mark size=1.3, only marks, mark=o, orange] table [col sep=comma, x=rm0, y=im0] {eigenvalues.csv};
      \addlegendentry{\tiny $m=0$}
      \addplot[mark size=1.5, only marks, mark=+, cyan] table [col sep=comma, x=rp1, y=ip1] {eigenvalues.csv};
      \addlegendentry{\tiny $m=m_{c}$}
      \addplot[mark size=1.5, only marks, mark=triangle, purple] table [col sep=comma, x=rp2, y=ip2] {eigenvalues.csv};
      \addlegendentry{\tiny $m=2m_{c}$}
    \end{axis}
  \end{tikzpicture}
  \caption{The first 240 eigenvalues at the critical parameter when $m_c=2$, $a=10$, $k=1$ and $N_y = 240$.
  \label{fig:eigs}
  }
\end{figure}
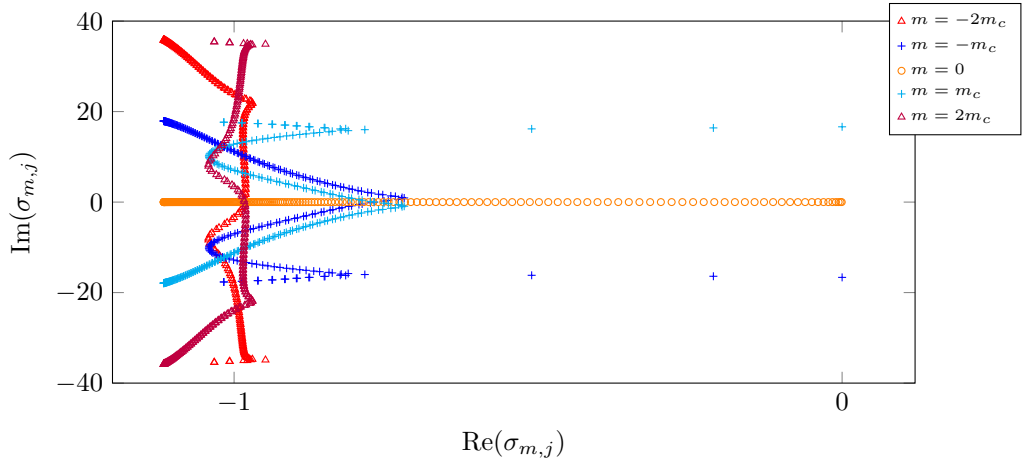

This eigenvalue problem is solved numerically by means of a standard Legendre-Galerkin method; see Appendix \ref{Sec_AppA}.
A typical picture of the spectrum near the criticality is given in~\autoref{fig:eigs}.
This figure shows that many eigenvalues are clustered near the imaginary axis at the
critical {parameter value $\Psi=\Psi_c$} as defined {in \eqref{Eq_Psic} below.}

We assume (as {confirmed numerically for the parameter regimes considered below}) that the eigenvalues are ordered so that for each {integer} $m$, $\sigma_{m, 1}$ has the largest real part among {the $\sigma_{m, j}$, for any nonnegative integer $j$}.

For each {(nonnegative) wavenumber} $m$, we define $\Psi_m$, {when} it exists, to be the value of $\Psi$ for which
the eigenvalue $\sigma_{m, 1}$ crosses the imaginary axis, that is
\begin{equation}\label{Pm}
  Re (\sigma_{m, 1}) = Re (\sigma_{-m, 1}) =
  \begin{cases}
    < 0, \text{ if } \Psi < \Psi_{m} \\
    = 0, \text{ if } \Psi = \Psi_{m} \\
    > 0, \text{ if } \Psi > \Psi_{m}
  \end{cases}.
\end{equation}

\begin{figure}[t]
  \centering
\begin{subfigure}
    \centering
    \begin{tikzpicture}[]
      \begin{axis}[x=3mm, y=100mm, ymin=0.05, ymax=.2, xmin=0, xmax=26, xlabel=$a$, ylabel=$\Psi$, title={$k=1$}]
        \addplot[thick, smooth, mark=*, mark size=.1mm]  table [col sep=comma, x=a, y=psi] {neutstab_k1_m1.csv};
        \addplot[thick, purple, mark=*, mark size=.1mm]  table [col sep=comma, x=a, y=psi] {neutstab_k1_m2.csv};
        \addplot[thick, red, mark=*, mark size=.1mm]  table [col sep=comma, x=a, y=psi] {neutstab_k1_m3.csv};
        \addplot[thick, brown, mark=*, mark size=.1mm]  table [col sep=comma, x=a, y=psi] {neutstab_k1_m4.csv};
        \node[] at (axis cs: 23.3, .16) {\small $\Psi_1$};%
        \node[purple] at (axis cs: 13.3, .16) {\small $\Psi_2$};;
        \node[red] at (axis cs: 9.1, .16) {\small $\Psi_3$};
        \node[brown] at (axis cs: 7, .16) {\small $\Psi_4$};
      \end{axis}
    \end{tikzpicture}
  \end{subfigure}
  \begin{subfigure}
    \centering
    \begin{tikzpicture}[]%
      \begin{axis}[x=3mm, y=300mm, ymin=0.03, ymax=.07, xmin=0, xmax=26, xlabel=$a$, ylabel=$\Psi$, title={$k=2$}]
        \addplot[thick, smooth, mark=*, mark size=.1mm]  table [col sep=comma, x=a, y=psi] {neutstab_k2_m1.csv};
        \addplot[thick, purple, mark=*, mark size=.1mm]  table [col sep=comma, x=a, y=psi] {neutstab_k2_m2.csv};
        \addplot[thick, red, mark=*, mark size=.1mm]  table [col sep=comma, x=a, y=psi] {neutstab_k2_m3.csv};
        \addplot[thick, brown, mark=*, mark size=.1mm]  table [col sep=comma, x=a, y=psi] {neutstab_k2_m4.csv};
        \node[] at (axis cs: 22.8, .06) {\small $\Psi_1$};%
        \node[purple] at (axis cs: 13.2, .06) {\small $\Psi_2$};;
        \node[red] at (axis cs: 9.1, .06) {\small $\Psi_3$};
        \node[brown] at (axis cs: 6.9, .06) {\small $\Psi_4$};
      \end{axis}
    \end{tikzpicture}
  \end{subfigure}
  \begin{subfigure}
    \centering
    \begin{tikzpicture}[]
      \begin{axis}[x=3mm, y=500mm, ymin=0.02, ymax=.04, xmin=0, xmax=26, xlabel=$a$, ylabel=$\Psi$, title={$k=3$}]
        \addplot[thick, smooth, mark=*, mark size=.1mm]  table [col sep=comma, x=a, y=psi] {neutstab_k3_m1.csv};
        \addplot[thick, purple, mark=*, mark size=.1mm]  table [col sep=comma, x=a, y=psi] {neutstab_k3_m2.csv};
        \addplot[thick, red, mark=*, mark size=.1mm]  table [col sep=comma, x=a, y=psi] {neutstab_k3_m3.csv};
        \addplot[thick, brown, mark=*, mark size=.1mm]  table [col sep=comma, x=a, y=psi] {neutstab_k3_m4.csv};
        \node[] at (axis cs: 22.6, .036) {\small $\Psi_1$};%
        \node[purple] at (axis cs: 13.5, .036) {\small $\Psi_2$};;
        \node[red] at (axis cs: 9.1, .036) {\small $\Psi_3$};
        \node[brown] at (axis cs: 6.9, .036) {\small $\Psi_4$};
      \end{axis}
    \end{tikzpicture}
  \end{subfigure}
  \caption{Neutral stability curves $\Psi_m, m=1,2,3,4$ defined in~\eqref{Pm}, for wind-stress
  profiles defined by $k=1, 2, 3$.   For values   of $\Psi > \Psi_m$, the shear flow  {$(\psi_1^s,\psi_2^s)$ in \eqref{eq:steady-state}}
  becomes unstable to a perturbation pattern with wavenumber $\alpha_m$.
  }
  \label{fig:neutstab}
\end{figure}
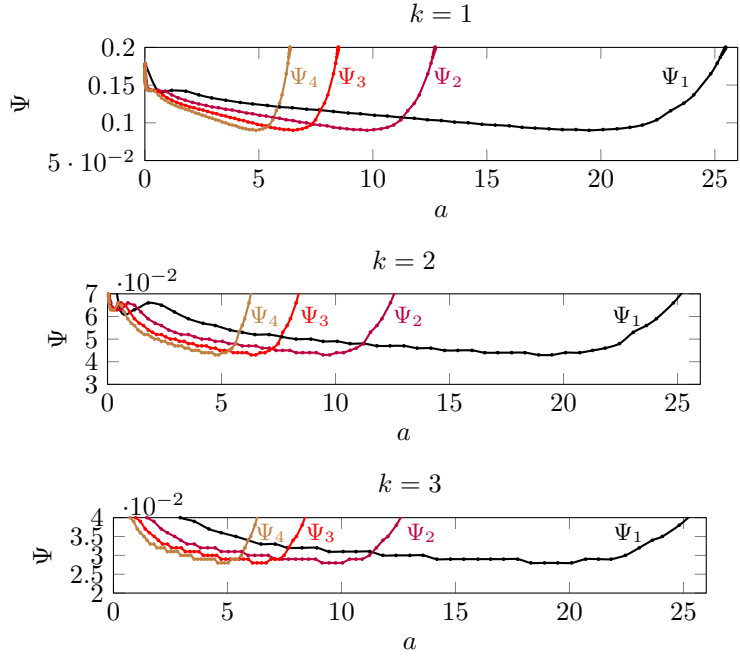

Hence $\Psi= \Psi_m$ defines a neutral stability curve in the {$(a,\Psi)$-plane}.
In \autoref{fig:neutstab}, these neutral curves are plotted for zonal wave
numbers $m=1,2,3,4$.

Our numerical analysis suggests that for {any} $m$, $\Psi_m$ {is well-defined}
only for aspect ratios of the basin characteristic lengths smaller than a threshold $a_m$ {(depending on $m$)}, that is for $a < a_m$. The
threshold $a_m$ is defined by a vertical asymptote condition, {namely}
\begin{displaymath}
  \lim_{a \to a_m-} \Psi_m = \infty.
\end{displaymath}
Moreover, {we numerically observe that the $a_m$ are ordered such that}
\begin{displaymath}
  \infty > a_1 > a_2 > \cdots
\end{displaymath}

We define then the {{\bf critical maximal shear flow's velocity}, $\Psi_c$,} and the critical zonal wavenumber, $m_c$, by
\begin{equation}\label{Eq_Psic}
  \Psi_c = \min_{m \in \mathbb{N}} \Psi_m,
\end{equation}
{and},
\begin{equation}\label{Eq_mc}
  m_c = \underset{m \in \mathbb{N}}{\mathrm{argmin}} \; \Psi_m,
\end{equation}
{respectively.}

{A} typical structure of the spectrum at the critical parameter {value}, $\Psi = \Psi_c$, is shown in \autoref{fig:eigs} where a {conjugate pair of} eigenvalues {is about to} cross the imaginary axis. The eigenvalues on the real axis {correspond to the} wavenumber $m=0$ and are always stable  {although they may be very close to zero as shown in \autoref{fig:eigs}.}

\begin{figure}[t]%
  \centering
  \includegraphics[height=0.5\textwidth, width=1\textwidth]{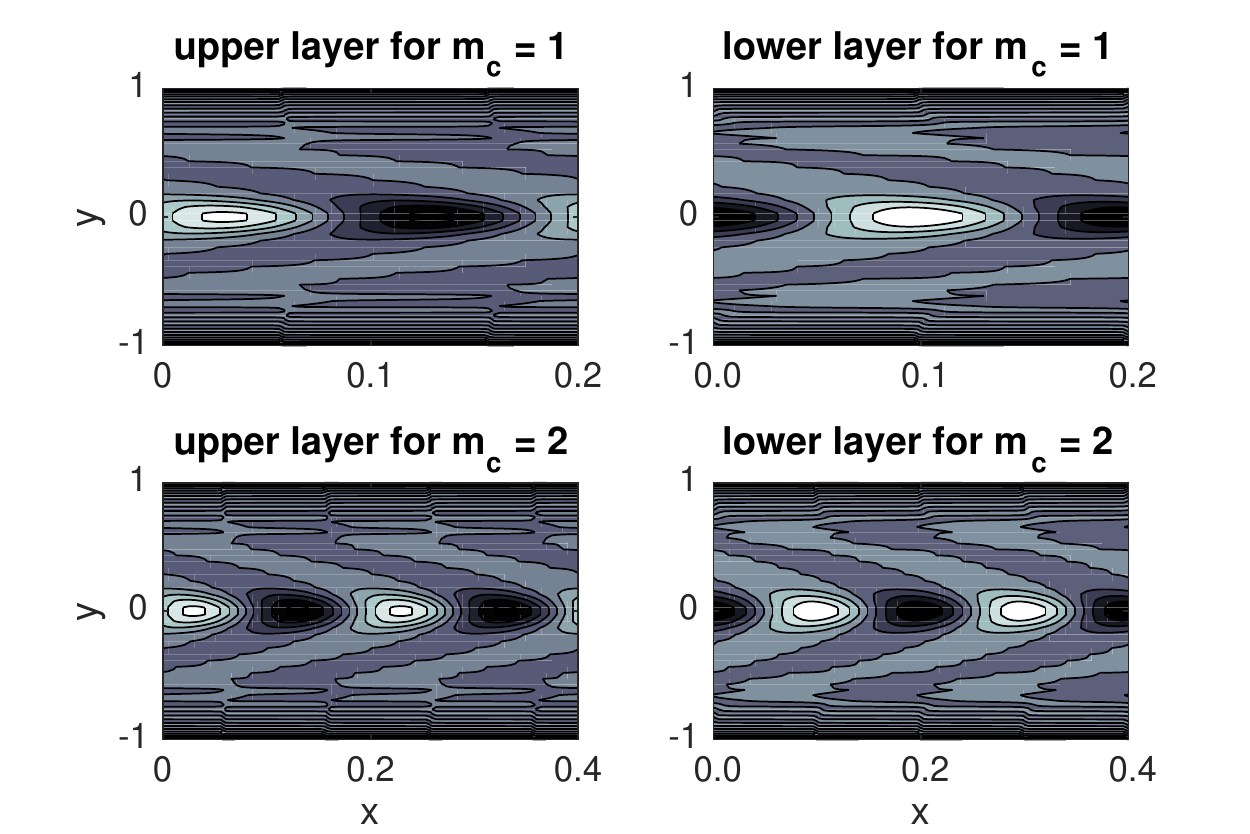}
  \caption{Real part of the upper and lower layers of the time-periodic
    solution $f_{m_c}$ (or equivalently of the dominant eigenmode
    $\psi_{m_c, 1}$) at $t=0$ for $k=1$ and $a = 10$ and $a =  5$,
    respectively.}
  \label{fig:crit_evec}
\end{figure}

To describe the solutions near the onset of transition $\Psi = \Psi_c$, we define the { spatio-temporal function}
\begin{equation} \label{f_m}
  f_m(x, y, t) = 2 \Re \left( e^{ i\Im(\sigma_{m, 1})t } \psi_{m, 1}(x, y) \right), \quad m\in \mathbb{Z},
\end{equation}
where $\sigma_{m, 1}$ is the first eigenvalue and $\psi_{m, 1}$ is its associated eigenfunction.
The spatial structure of the eigenmodes $\psi_{m_c, 1}$ { is} shown in~\autoref{fig:crit_evec}, revealing the well-known ``banana-shaped'' patterns characteristics of baroclinic instability.

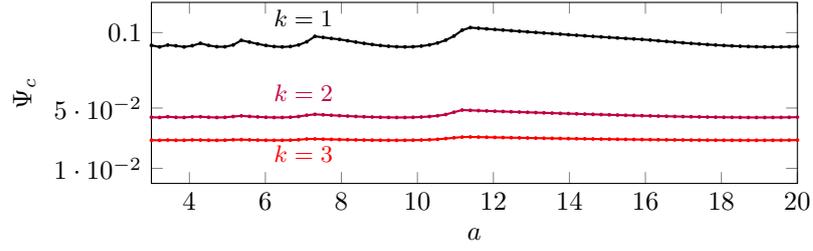
\begin{figure}[t]
  \centering
  \begin{tikzpicture}[]
    \begin{axis}[x=5mm, y=200mm, xmin = 3, xmax=20, ymin=0.0, ymax=.12, xlabel=$a$, ylabel=$\Psi_{c}$, ytick ={1e-2, 5e-2, 1e-1, 5e-1}]
      \addplot[black, thick, mark size=.3, mark=*] table [col sep=comma, x=a, y=psi] {transno_k1.csv};
      \addplot[purple, thick, mark size=.3, mark=*] table [col sep=comma, x=a, y=psi] {transno_k2.csv};
      \addplot[red, thick, mark size=.3, mark=*] table [col sep=comma, x=a, y=psi] {transno_k3.csv};
      \node[] at (axis cs: 7, .11) {\small $k=1$};
      \node[purple] at (axis cs: 7, .06) {\small $k=2$};
      \node[red] at (axis cs: 7, .02) {\small $k=3$};
    \end{axis}
  \end{tikzpicture}
  \caption{The {critical maximal shear flow's velocity} $\Psi_{c}$ {dependence on} the channel's aspect ratio $a$ for $k=1, 2$ and  $k=3$.\label{fig:critpsi}}
\end{figure}

The values of $\Psi_c$ with respect to the aspect ratio $a$ for $k=1,2,3$ is shown in~\autoref{fig:critpsi}.
By the previous remarks,
\begin{equation}\label{ls}
  \lim_{a \to a_1-} \Psi_c = \lim_{a \to a_1-} \Psi_1 = \infty.
\end{equation}
By~\eqref{ls}, \emph{for $a>a_{1}$, the system is linearly stable.}
As is expected, the neutral stability curves (\autoref{fig:neutstab}) approach
the asymptote $\Psi_{c} \to \infty$ as $a$ {converges} to the critical aspect ratio
$a_1$ over which the system is linearly stable for all $\Psi$.
{Our numerical results in Figure 4 show that for aspect ratios $a$ in the range $3 \le a \le 20$, the critical maximal shear velocity $\Psi_c \approx 0.1, 0.04, 0.03$ for $k=1, 2, 3$ respectively.}
By \eqref{eq:1}, this value of critical maximal shear velocity corresponds to an upper layer friction  that is approximately,
\begin{displaymath}
  r_1 \approx \frac{ \tau \beta }{ k^{2} \pi^{2} \Psi_{c} } \approx
  \frac{{10^3}}{ k },
\end{displaymath}
which is indeed much larger than can be justified from dissipative processes in the ocean
but, as explained in \autoref{Sec2}, is needed here {to connect the background zonal jet to the applied wind stress as a Sverdrup balance is absent in the original model formulation.}


 The friction term in the lower layer however is physical (Ekman friction) and for this study, it is fixed at $r_2 = 5.0$, see~\autoref{tab:defp}.
Also, from \autoref{fig:neutstab}, we see that for small aspect ratios, many modes become unstable as $\Psi_{c}$ is exceeded.

For $a < a_1$, the system has a first transition at $\Psi = \Psi_c$ and exactly one of the following two principal of exchange of stability (PES) condition holds:
\begin{equation}\label{PES}
  \begin{aligned}
    & Re (\sigma_{m, 1}) = Re (\sigma_{-m, 1})
    \begin{cases}
      < 0, \text{ if } \Psi < \Psi_{c} \\
      = 0, \text{ if } \Psi = \Psi_{c} \\
      > 0, \text{ if } \Psi > \Psi_{c}
    \end{cases} && \text{ if } m=m_c \\
    & Re (\sigma_{m, 1}) = Re (\sigma_{-m, 1})  < 0 && \text{ if } m \ne m_c
  \end{aligned}
\end{equation}

\begin{equation}\label{PES2}
  \begin{aligned}
    & Re (\sigma_{m, 1}) = Re (\sigma_{-m, 1})
    \begin{cases}
      < 0, \text{ if } \Psi < \Psi_{c} \\
      = 0, \text{ if } \Psi = \Psi_{c} \\
      > 0, \text{ if } \Psi > \Psi_{c}
    \end{cases} && \text{ if } m \in \{ m_c, m_c+1 \} \\
    & Re (\sigma_{m, 1}) = Re (\sigma_{-m, 1}) < 0 && \text{ if } m \notin \{ m_c, m_c+1 \}.
  \end{aligned}
\end{equation}

According to \eqref{PES} and \eqref{PES2} either, two or four eigenvalues become unstable as $\Psi$ crosses $\Psi_c$.
The case of two critical eigenvalues is generic and results into a Hopf bifurcation.
The case of four critical eigenvalues results in a double Hopf bifurcation and requires the fine-tuning of the aspect ratio $a = a_{DH}$ so that $\Psi_c = \Psi_{m_c} = \Psi_{m_c+1}$. The values of $a_{DH}$ where double Hopf transition occurs is given in~\autoref{tab:doublehopf}.

\begin{table}[tb]
  \centering
  \begin{tabular}{|c|c|c|c|c|c|c|c|c|c|c|}
    \hline
    $k$ & $m_c$ & $a_{DH}$ & $\Psi_c$ & $Re(A)$ & $Re(B)$ & $Re(C)$ & $Re(D)$ & $\delta$ & $\theta$ & $\delta \theta$ \\ \hline
    1   & 1     & 11.26    & 0.10     & -0.25   & -1.00   & 0.01    & -0.33   & -0.05    & 3.08                          & -0.15                                \\ \hline
    1   & 2     & 7.30     & 0.10     & -0.37   & -1.00   & -0.13   & -0.40   & 0.36     & 2.51                          & 0.89                                 \\ \hline
    1   & 3     & 5.38     & 0.10     & -0.95   & -1.00   & -0.22   & -0.61   & 0.23     & 1.65                          & 0.37                                 \\ \hline
    1   & 4     & 4.23     & 0.09     & -0.81   & -1.00   & -0.29   & -0.61   & 0.36     & 1.63                          & 0.58                                 \\ \hline
    2   & 1     & 11.18    & 0.05     & -0.15   & -1.00   & 0.12    & -0.23   & -0.77    & 4.39                          & -3.38                                \\ \hline
    2   & 2     & 7.20     & 0.05     & -0.34   & -1.00   & -0.04   & -0.46   & 0.13     & 2.20                          & 0.28                                 \\ \hline
    2   & 3     & 5.27     & 0.05     & -0.37   & -1.00   & -0.14   & -0.45   & 0.38     & 2.24                          & 0.86                                 \\ \hline
    2   & 4     & 4.14     & 0.05     & -0.38   & -1.00   & -0.21   & -0.44   & 0.55     & 2.25                          & 1.25                                 \\ \hline
  \end{tabular}
  \caption{The double Hopf transition numbers. Here $k=1,2$ is the wavenumber of the forcing, $m_c$ and $m_c+1$ are the wavenumbers of the first two critical modes which become unstable simultaneously at the critical aspect ratio $a_{DH}$. $A$, $B$, $C$, $D$ are the coefficients of the double Hopf transition normalized with respect to the maximum absolute value of those coefficients.
    The parameters $\delta$ and $\theta$ are defined by \eqref{delta}.
    ~\label{tab:doublehopf}}
\end{table}

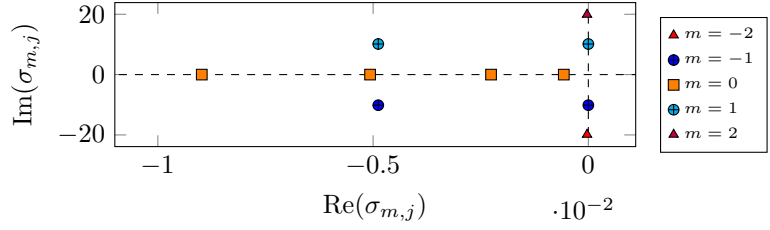
\begin{figure}[h]%
  \centering
  \begin{tikzpicture}[]
    \begin{axis}[xmin=-0.011, xtick={-0.01,-0.005,0}, y=.4mm,
      xlabel=$\Re(\sigma_{m, j})$, ylabel=$\Im(\sigma_{m, j})$, legend
      style={at={(1.15,0.9)},anchor=north,legend cell align=left}]
      \addplot[mark size=2, only marks, mark=triangle*, draw=black, fill=red] table [col sep=comma, x=rm2, y=im2] {eigenvalues_dh.csv};
      \addlegendentry{\tiny $m=-2$}
      \addplot[mark size=2, only marks, mark=oplus*, draw=black, fill=blue] table [col sep=comma, x=rm1, y=im1] {eigenvalues_dh.csv};
      \addlegendentry{\tiny $m=-1$}
      \addplot[mark size=2, only marks, mark=square*, draw=black, fill=orange] table [col sep=comma, x=rm0, y=im0] {eigenvalues_dh.csv};
      \addlegendentry{\tiny $m=0$}
      \addplot[mark size=2, only marks, mark=oplus*, draw=black, fill=cyan] table [col sep=comma, x=rp1, y=ip1] {eigenvalues_dh.csv};
      \addlegendentry{\tiny $m=1$}
      \addplot[mark size=2, only marks, mark=triangle*, draw=black, fill=purple] table [col sep=comma, x=rp2, y=ip2] {eigenvalues_dh.csv};
      \addlegendentry{\tiny $m=2$}
      \draw[dashed] (0,-20) -- (0,20);
      \draw[dashed] (-0.015,0) -- (0.1,0);
    \end{axis}
  \end{tikzpicture}
  \caption{The spectrum near the double Hopf aspect ratio $a_{DH} = 11.263$ for $k=1$. The first four critical eigenvalues $\sigma_{m,1}$, $m$ in $\left\{ -2,-1,1,2 \right\}$, can be seen on the imaginary axis.}
  \label{fig:doubleeigs}
\end{figure}

Although the double Hopf transition is not generic, its analysis gives an insight into {regimes of moderate values of $\Psi$} where multiple eigenvalues are unstable.
In \autoref{fig:doubleeigs}, we {show} the dominant part of the spectrum of the linearized operator at a critical aspect ratio $a_{DH}$.

{We recall that the neutral stability curves (shown in \autoref{fig:neutstab}) are found by identifying the set of parameter values (here in the ($a$,$\Psi$)-plane) for which the real part of the critical eigenvalue is zero while the rest of the eigenvalues have a negative real part, i.e.~are associated with a stable mode. By a continuity argument, we can thus infer that in a neighborhood of such critical curves the PES condition is satisfied, for instance when $\Psi$ crosses the critical value $\Psi_c$ (depending on $a$ and $k$) shown in Figure 2.}

The PES condition~\eqref{PES} has been rigorously verified for Kolmogorov flows in~\cite{meshalkin1961} via a continued fraction method.
This method has later been extended for the single layer QG model for the $k=1$ case in~\cite{simonnet2003} and for $k\ge 2$ in~\cite{lu2020} where $k$ is the forcing frequency in~\eqref{eq:main-1}. It is still an open problem to {rigorously verify the PES condition} for the current problem.

\subsection{Hopf Bifurcation}\label{Sec_Hopf}
We first investigate the generic Hopf transition scenario based on the attractor bifurcation theorem \cite[Theorem 5.2]{ma2005} and  the dynamical transition theorem \cite[Theorem 2.1.3]{ptd}. For proofs of the following lemma and theorem, see Appendix \ref{Sec_AppB}).
\begin{lemma}\label{lemma1}
  Assume that the PES condition~\eqref{PES} holds.
  Then the transition and stability of the steady state solution \eqref{eq:steady-state} {to Eq.~}\eqref{eq:main-abstract-form}, in the vicinity of the critical maximal shear  {flow's velocity} $\Psi = \Psi_c$ and for any sufficiently small initial {data}, are equivalent to the stability of the zero solution of the {following reduced} equation
  \begin{equation}\label{req-m1}
    \frac{dz}{dt} = \sigma_{m_{c}, 1} z + P z \abs{z}^{2}+o(\abs{z}^{3}),
  \end{equation}
where $z(t)$ {denotes the complex} amplitude {aiming at approximating the projection of the model's solution
  onto the critical mode $\psi_{m_{c}, 1}$. The coefficient $P$ is defined in Eq.~\eqref{Eq_P} of Appendix \ref{Sec_AppB} and is called the \textbf{transition number}.}
\end{lemma}
The analysis of \autoref{lemma1} yields the following theorem.
\begin{theorem}\label{thm1}
  Assume that the first critical eigenvalue is {purely complex with  a simple multiplicity}, that is the PES condition~\eqref{PES} is satisfied.
  Then the following assertions hold true.
\begin{enumerate}
  \item[(1)] If $\Re(P)<0$, then {Eq.~\eqref{eq:main-abstract-form} in $H_1$ (see \eqref{Func_spaces})} undergoes a continuous transition accompanied by a supercritical Hopf bifurcation on $\Psi > \Psi_c$, {with $\Psi_c$ defined in \eqref{Eq_Psic}}.
  In particular, the steady-state solution bifurcates to a stable periodic solution $\psi$ on $\Psi>\Psi_c$, satisfying $\psi \to \mathbf{0}$ as $ \Psi \to \Psi_c$ and has the following approximation
  \begin{equation} \label{thm1 solution}
    \psi(x, y, t) = \left( \frac{-\Re(\sigma_{m_{c}, 1}) }{\Re(P)}\right)^{\frac{1}{2}}
    f_{m_{c}}(x, y, t)+ o \left( \abs{ \Psi - \Psi_c}^{\frac{1}{2}} \right).
  \end{equation}
  The spatial structure of the time periodic solution $\psi$ is shown at $t=0$ for different aspect ratios $a$ in~\autoref{fig:crit_evec}.
  \item[(2)] If $\Re(P)>0$, then {Eq.~\eqref{eq:main-abstract-form} in $H_1$ (see \eqref{Func_spaces})} undergoes a jump transition on $\Psi<\Psi_c$ accompanied by a subcritical Hopf bifurcation.
  In particular, an unstable periodic {solution} $\psi$ given by~\eqref{thm1 solution} bifurcates on $\Psi < \Psi_c$ and there is no periodic solution bifurcating from $\mathbf{0}$ on $\Psi>\Psi_c$. Moreover, there is a singularity separation at some $\Psi_{s} < \Psi_c$ generating an attractor and an unstable periodic {solution} $\psi$.
\end{enumerate}
\end{theorem}

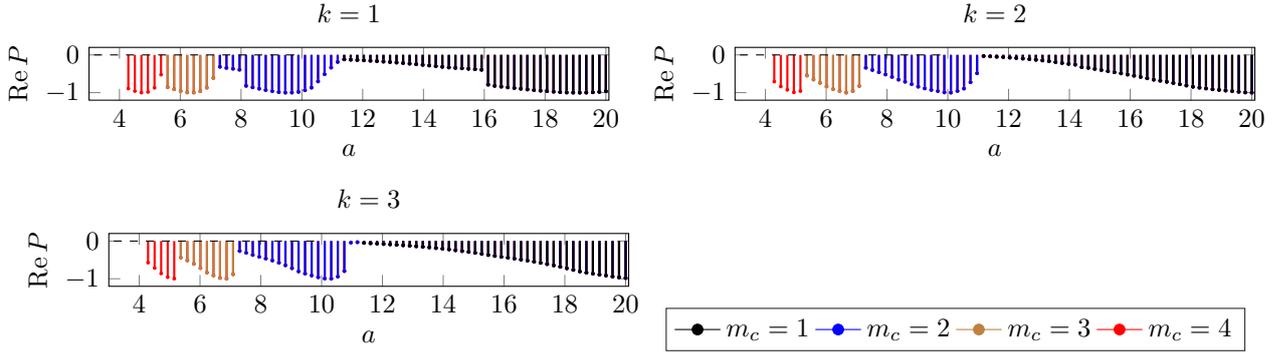
\begin{figure}[tb]
  \centering
  \begin{subfigure}
    \centering%
    \begin{tikzpicture}
      \begin{axis}[ycomb, xmin=3, xmax=20.1, x=4mm, y=5mm, ymin = -1.2, ymax= 0.2, xlabel=$a$, ylabel=$\Re P$, ytick={-1, 0}, title={$k=1$}]

        \addplot+[red, thick, mark size=.3, mark=*] table [col sep=comma, x=a,
        y expr={\thisrow{m} < 5 ? \thisrow{P} : NaN},] {transno_k1.csv};

        \addplot+[brown, thick, mark size=.3, mark=*] table [col sep=comma, x=a,
        y expr={\thisrow{m} < 4 ? \thisrow{P} : NaN},] {transno_k1.csv};

        \addplot+[blue, thick, mark size=.3, mark=*] table [col sep=comma, x=a,
        y expr={\thisrow{m} < 3 ? \thisrow{P} : NaN},] {transno_k1.csv};

        \addplot+[black, thick, mark size=.3, mark=*] table [col sep=comma, x=a,
        y expr={\thisrow{m} < 2 ? \thisrow{P} : NaN},] {transno_k1.csv};
        \draw[dashed] (0, 0) -- (10, 0);
      \end{axis}
    \end{tikzpicture}
  \end{subfigure}
  \begin{subfigure}
    \centering
    \begin{tikzpicture}
      \begin{axis}[ycomb, xmin=3, xmax=20.1, x=4mm, y=5mm, ymin = -1.2, ymax= 0.2,  xlabel=$a$, ylabel=$\Re P$, ytick={-1, 0}, title={$k=2$}]

        \addplot+[red, thick, mark size=.3, mark=*] table [col sep=comma, x=a,
        y expr={\thisrow{m} < 5 ? \thisrow{P} : NaN},] {transno_k2.csv};

        \addplot+[brown, thick, mark size=.3, mark=*] table [col sep=comma, x=a,
        y expr={\thisrow{m} < 4 ? \thisrow{P} : NaN},] {transno_k2.csv};

        \addplot+[blue, thick, mark size=.3, mark=*] table [col sep=comma, x=a,
        y expr={\thisrow{m} < 3 ? \thisrow{P} : NaN},] {transno_k2.csv};

        \addplot+[black, thick, mark size=.3, mark=*] table [col sep=comma, x=a,
        y expr={\thisrow{m} < 2 ? \thisrow{P} : NaN},] {transno_k2.csv};
        \draw[dashed] (0, 0) -- (10, 0);
      \end{axis}%
    \end{tikzpicture}
  \end{subfigure}
  \begin{subfigure}
    \centering
    \begin{tikzpicture}
      \begin{axis}[ycomb, xmin=3, xmax=20.1, x=4mm, y=5mm, ymin = -1.2, ymax= 0.2,  xlabel=$a$, ylabel=$\Re P$, ytick={-1, 0}, title={$k=3$}]

        \addplot+[red, thick, mark size=.3, mark=*] table [col sep=comma, x=a,
        y expr={\thisrow{m} < 5 ? \thisrow{P} : NaN},] {transno_k3.csv};

        \addplot+[brown, thick, mark size=.3, mark=*] table [col sep=comma, x=a,
        y expr={\thisrow{m} < 4 ? \thisrow{P} : NaN},] {transno_k3.csv};

        \addplot+[blue, thick, mark size=.3, mark=*] table [col sep=comma, x=a,
        y expr={\thisrow{m} < 3 ? \thisrow{P} : NaN},] {transno_k3.csv};

        \addplot+[black, thick, mark size=.3, mark=*] table [col sep=comma, x=a,
        y expr={\thisrow{m} < 2 ? \thisrow{P} : NaN},] {transno_k3.csv};
        \draw[dashed] (0, 0) -- (10, 0);
      \end{axis}
    \end{tikzpicture}
  \end{subfigure}
  \begin{subfigure}
    \centering
    \begin{tikzpicture}
      \begin{axis}[%
        hide axis,
        xmin=20,
        xmax=40,
        ymin=0,
        ymax=0.3,
        legend style={draw=white!15!black,legend cell align=center, legend columns=4}
        ]
        \addlegendimage{black, mark=*}
        \addlegendentry{$m_c=1$};
        \addlegendimage{blue, mark=*}
        \addlegendentry{$m_c=2$};
        \addlegendimage{brown, mark=*}
        \addlegendentry{$m_c=3$};
        \addlegendimage{red, mark=*}
        \addlegendentry{$m_c=4$};
      \end{axis}
    \end{tikzpicture}
  \end{subfigure}
  \caption{The real part of the transition number $P$ compared to the channel aspect ratio $a$ normalized by the largest absolute value of $\Re P$. The parameters are as set in \autoref{tab:defp}.}
  \label{figP}
\end{figure}

When the PES condition~\eqref{PES} holds, the system exhibits a Hopf bifurcation as described by~\autoref{thm1}.
The type of transition boils down to the { determination of the transition number $P$ given in \eqref{Eq_P} of Appendix \ref{Sec_AppB}.}
For the practical calculation of this number, we refer to Appendix \ref{Sec_AppC}.  {A numerical evaluation of $P$ as reported} in~\autoref{figP} shows that for {a low-frequency forcing ($k=1, 2, 3$)}, \emph{generally} $Re(P) < 0$ and as a result \emph{only continuous transition (supercritical Hopf bifurcation) is possible for the parameter regime we have selected}.
In \autoref{figP}, we also display the critical wavenumber $m_c$. In the range $4 \le a \le 20$, the critical wavenumber $m_c$ is found {to range from $1$ to $4$.}

There are discontinuities in $P$ vs $a$ plot in~\autoref{figP} of the transition number which are due to the changes in the critical zonal wavenumber $m_c$. These discontinuities take place at double Hopf bifurcation aspect ratios where two consecutive zonal wavenumbers become critical simultaneously which is investigated in the next section.
However, there are also discontinuities in~\autoref{figP}, $k=1$ case (for example near $a=16$) whose origin is mysterious.

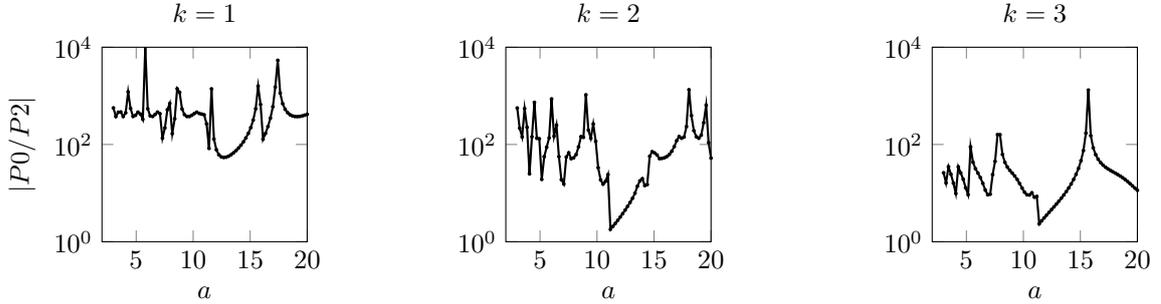
\begin{figure}[tb]%
  \begin{minipage}[t]{0.33\textwidth}%
    \centering
    \begin{tikzpicture}[]
      \begin{axis}[ymode=log, x=1.5mm, y=2.8mm, xmin = 2, xmax=20, ymin=1E0,
        ymax=1E4, xlabel=$a$, ylabel=$\lvert P0/P2 \rvert$, title={$k=1$}]
        \addplot[black, thick, mark size=.3, mark=*] table [col sep=comma, x=a,
        y expr= {abs( \thisrowno{1} / \thisrowno{2})}] {transno_k1.csv};
      \end{axis}
    \end{tikzpicture}
  \end{minipage}%
  \begin{minipage}[t]{0.33\textwidth}
    \centering
    \begin{tikzpicture}[]
      \begin{axis}[ymode=log, x=1.5mm, y=2.8mm, xmin = 2, xmax=20, ymin=1E0,
        ymax=1E4, xlabel=$a$, title={$k=2$}]
        \addplot[black, thick, mark size=.3, mark=*] table [col sep=comma, x=a,
        y expr= {abs( \thisrowno{1} / \thisrowno{2})}] {transno_k2.csv};
      \end{axis}
    \end{tikzpicture}
  \end{minipage}%
  \begin{minipage}[t]{0.33\textwidth}
    \centering
    \begin{tikzpicture}[]
      \begin{axis}[ymode=log, x=1.5mm, y=2.8mm, xmin = 2, xmax=20, ymin=1E0,
        ymax=1E4, xlabel=$a$, title={$k=3$}]
        \addplot[black, thick, mark size=.3, mark=*] table [col sep=comma, x=a,
        y expr= {abs( \thisrowno{1} / \thisrowno{2})}] {transno_k3.csv};
      \end{axis}%
    \end{tikzpicture}
  \end{minipage}
  \caption{The effect of $P_{0}$ (of $m=0$ modes) compared to the effect $P_{2}$
    (of $2m_{c}$ modes) on the transition number.\label{fig: P0-P2}}
\end{figure}

As detailed in Appendix \ref{Sec_AppB}, the transition number $P$ accounts for two types of nonlinear interactions between the eigenmodes, and is written
\begin{displaymath}
  P = P_{0} + P_{2},
\end{displaymath}
where $P_{0}$ accounts for nonlinear interactions between the critical modes and the zonally homogeneous {(stable)}  modes $m=0$ (see \eqref{Eq_P0} below), while $P_{2}$ accounts for the interactions between the critical modes and the modes having a wavenumber twice that of the critical modes (see \eqref{Eq_P2} below).
A  comparison of  typical numerical values of $P_0$ and $P_2$ shows that $P_{2}$ is several orders of magnitudes smaller than $P_{0}$; see \autoref{fig: P0-P2}. We refer to \cite[Theorem III.1]{chekroun2020efficient} for a {similar} transition number diagnosing also the type of Hopf bifurcations arising, generically,  in delay differential equations, and whose nature is also characterized by the interactions of linearized modes through the model’s non-linear terms.

\begin{figure}[t]
\centering
  \begin{subfigure}
    \centering
    \begin{tikzpicture}[]
      \begin{axis}[x=4mm, xmin = -3, xmax=14, y=40mm, xlabel=$a$, title={$k=1$}]
        \addplot[mark size=.3] table [col sep=comma,  x=Var1, y=Var3]  {P0transno_k1.csv} node[left,pos=0] {\tiny $j=2$};
        \addplot+[mark size=.3] table [col sep=comma, x=Var1, y=Var5]  {P0transno_k1.csv} node[right,pos=1] {\tiny $j=4$};
        \addplot+[mark size=.3] table [col sep=comma, x=Var1, y=Var7]  {P0transno_k1.csv} node[left,pos=0] {\tiny $j=6$};
        \addplot+[mark size=.3] table [col sep=comma, x=Var1, y=Var9]  {P0transno_k1.csv} node[right,pos=1] {\tiny $j=8$};
        \addplot+[mark size=.3] table [col sep=comma, x=Var1, y=Var11] {P0transno_k1.csv} node[left,pos=0] {\tiny $j=10$};
      \end{axis}
    \end{tikzpicture}
  \end{subfigure}
  \begin{subfigure}
    \centering%
    \begin{tikzpicture}[]
      \begin{axis}[x=4mm, xmin = -3, xmax=14, y=120mm, xlabel=$a$, title={$k=2$}]
        \addplot[mark size=.3] table [col sep=comma,  x=Var1, y=Var3]  {P0transno_k2.csv} node[left,pos=0] {\tiny $j=2$};
        \addplot+[mark size=.3] table [col sep=comma, x=Var1, y=Var5]  {P0transno_k2.csv} node[right,pos=1] {\tiny $j=4$};
        \addplot+[mark size=.3] table [col sep=comma, x=Var1, y=Var7]  {P0transno_k2.csv} node[left,pos=0] {\tiny $j=6$};
        \addplot+[mark size=.3] table [col sep=comma, x=Var1, y=Var9]  {P0transno_k2.csv} node[right,pos=1] {\tiny $j=8$};
        \addplot+[mark size=.3] table [col sep=comma, x=Var1, y=Var11] {P0transno_k2.csv} node[left,pos=0] {\tiny $j=10$};
      \end{axis}
    \end{tikzpicture}
  \end{subfigure}
  \caption{$P_{0,j} / \abs{ P_{0,2} }$ for even values of $j$.
    $P_{0,j}=0$ for odd values of $j$.
    \label{fig:P0j}}
\end{figure}
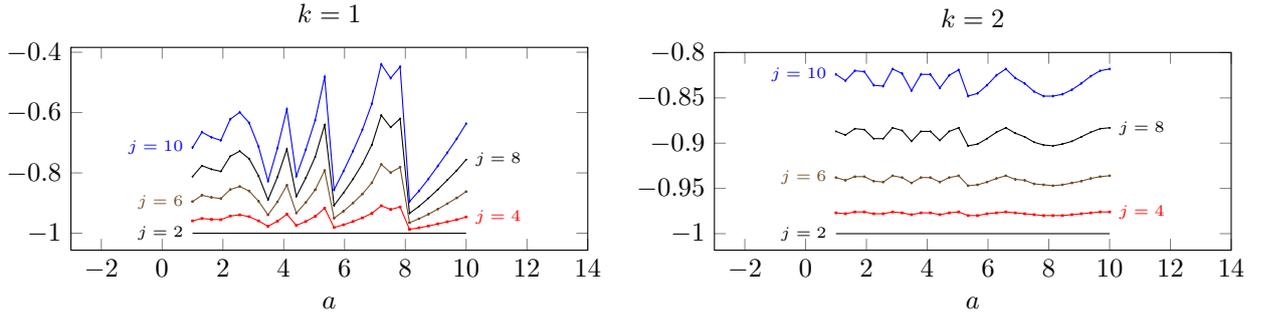

{In our case,}
\begin{displaymath}
  P_0 = \sum_{j=0}^{\infty} P_{0, j}
\end{displaymath}
where $P_{0, j}$ measures the contribution {to $P$} of the $j$-th mode {with zero-wavenumber ($m=0$) interacting with the critical mode}; {see  \eqref{Eq_P0} again}.
\autoref{fig:P0j} shows that the {dependence of $P_{0,j}$ on} $j$ is essentially linear.
We believe the results in \autoref{fig: P0-P2} and \autoref{fig:P0j} may help when choosing the modes to include in a simulation when the maximal shear velocity is well above the criticality.

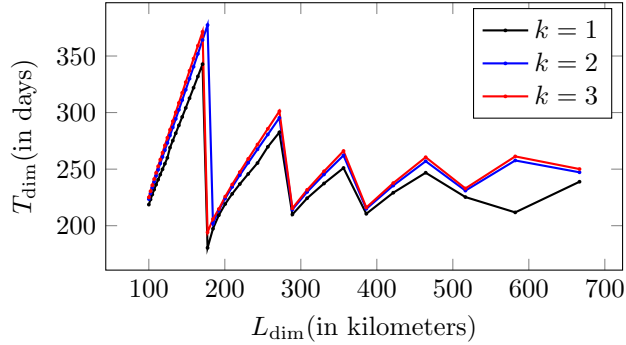
\begin{figure}[tb]
  \centering
    \begin{tikzpicture}[]
      \begin{axis}[x=.10mm, y=0.15mm, xlabel=$L_{\text{dim}}$(in kilometers), ytick = {100, 150, 200,250,300,350}, ylabel=$T_{\text{dim}}$(in days)]
        \addplot[black, thick, mark size=.3, mark=*] table [col sep=comma, x=L, y=T] {period_k1.csv};
        \addplot[blue, thick, mark size=.3, mark=*] table [col sep=comma, x=L, y=T] {period_k2.csv};
        \addplot[red, thick, mark size=.3, mark=*] table [col sep=comma, x=L, y=T] {period_k3.csv};
        \legend{$k=1$,$k=2$,$k=3$}
      \end{axis}
    \end{tikzpicture}
  \caption{The (dimensional) time period $T_{\text{dim}} = \frac{2\pi}{\Im(\sigma_{m_c,1})} \frac{L_y}{U}$ compared to the (dimensional) length of the channel $L_{\text{dim}} = 2L_y/a$ of the both stable and unstable bifurcated time periodic solution \eqref{thm1 solution} in the Hopf bifurcation case. Here $L_y$ and $U$ are the characteristic scales defined in \autoref{tab:defp}. The jumps in the derivative of the time period of the bifurcated solution is due to the change of the imaginary part $\Im \sigma_{m_c,1}$ of the critical eigenvalue at the double Hopf aspect ratios $a_{DH}$.
  \label{fig:per}}
\end{figure}

We also compare the dimensional time period of the bifurcated solution \eqref{thm1 solution} to the (dimensional) length of the channel in \autoref{fig:per}. With the default parameters as chosen in \autoref{tab:defp}, our simulations yield a solution with time period of 180-380 days depending on the channel length of 100-700kms.

\subsection{Double Hopf Bifurcation}\label{Sec_doubleHopf}
In this section we are interested in the transitions that take place at the critical aspect ratios $a_{DH}$ where four modes with consecutive wavenumbers $m_c$, $m_c+1$ become unstable as given by the PES
condition~\eqref{PES2}.

We first present the reduced equations in this case (for proofs, see Appendix \ref{Sec_AppD}).
\begin{lemma}\label{lemma2}
  Assume that the PES condition \eqref{PES2} holds.
  Then, the transition and stability of the steady state solution \eqref{eq:steady-state} of {Eq.~\eqref{eq:main-abstract-form} in $H_1$ (see \eqref{Func_spaces})}, in the vicinity of {the critical zonal shear flow's velocity}, $\Psi = \Psi_c$, and for any sufficiently small initial {data} are equivalent to the stability of the zero solution of the following {system}
  \begin{equation}\label{req-m2}
    \begin{aligned}
      \frac{d z_{1}}{d t} & = \sigma_{m_c, 1} z_{1}
      + z_{1} (A \abs{z_{1}}^{2} + B \abs{z_{2}}^{2}) + o( \abs{ z })^{3}),\\
      \frac{dz_{2}}{d t} & = \sigma_{m_c+1, 1} z_{2}
      + z_{2}(C \abs{z_{1}}^{2} + D \abs{z_{2}}^{2}) + o( \abs{ z })^{3}),
    \end{aligned}
  \end{equation}
  where $z_1(t)$ {and} $z_2(t)$ denote the {complex} amplitudes {aiming at approximating the projection of the model's solution  onto the critical modes $\psi_{m_{c}, 1}$ and $\psi_{m_{c}+1, 1}$, respectively.}

The \textbf{transition numbers} $A, B, C, D$ are determined by the nonlinear interactions of {these critical modes} with higher modes given by~\eqref{ABCD}.
More precisely, the terms $A$ and $D$ account for the self-interactions among the critical modes, while the terms $B$ and $C$ account for the cross-interactions between the critical modes with the higher modes.
\end{lemma}

It is known that the equation~\eqref{req-m2} exhibit a zoo of dynamical behaviors.
We refer to~\cite{kuznetsov2004} for a detailed analysis of all possible cases.
Here, we restrict our attention to the case
\begin{equation} \label{theta-delta-condition}
  \Re(A)< 0, \, \Re(B) < 0, \Re(B) + \Re(C) < 0, \Re(D)<0,
\end{equation}
which is the only case we observe in our numerical experiments, see~\autoref{tab:doublehopf}.
Under these conditions it is known that the transition is continuous (see Theorem 2.3 in~\cite{li2020b}).
For the next theorem, let us define the numbers
\begin{equation} \label{lambda}
  \lambda_1 = Re(\sigma_{m_c, 1}), \qquad
  \lambda_2 = Re(\sigma_{m_c+1, 1}).
\end{equation}
\begin{equation}\label{delta}
  \delta = \frac{ Re(C) }{ Re(A) }, \qquad
  \theta = \frac{ Re(B)  }{ Re(D) }
\end{equation}
\begin{displaymath}
  \eta_1 = \left( \frac{\lambda_1 - \theta \lambda_2 }{(\theta \delta - 1) \Re(A)}\right)^{\frac{1}{2}}, \qquad
  \eta_2 = \left( \frac{\lambda_2 - \delta \lambda_1   }{(\theta \delta - 1) \Re(D)}\right)^{\frac{1}{2}}.
\end{displaymath}
{ Recalling $f_{m_c}$ defined in \eqref{f_m} (with $m=m_c$), we define the following spatio-temporal profiles}
\begin{equation}\label{per-qper}
  \begin{aligned}
    &\psi_{p}^{m_c}(x, y, t) = \left( - \frac{\lambda_1 }{\Re(A)}\right)^{\frac{1}{2}} f_{m_c}(x, y, t) + o \left( \abs{ \Psi - \Psi_c}^{\frac{1}{2}} \right),\\
    &\psi_{p}^{m_c+1}(x, y, t) = \left( \frac{\lambda_2 }{\Re(D)}\right)^{\frac{1}{2}} f_{m_c+1}(x, y, t) + o \left( \abs{ \Psi - \Psi_c}^{\frac{1}{2}} \right),\\
    & \psi_{qp}(x, y, t) = \eta_1 f_{m_{c}}(x, y, t) + \eta_2 f_{m_c+1}(x, y, t) +
    o \left( \abs{ \Psi - \Psi_c}^{\frac{1}{2}} \right)\\
\end{aligned}
\end{equation}

\begin{figure}[tb]
\centering
\subfigure[$\theta>0$, $\delta>0$, $\theta \delta > 1$]{
\begin{tikzpicture}[scale = 2 ]
  \draw[thick, dashed] (0, 0) -- (0:1.2);
  \draw[thick, dashed] (0, 0) -- (90:1.2);
  \draw[thick, dashed] (0, 0) -- (180:.5);
  \draw[thick, dashed] (0, 0) -- (270:.5);
  \foreach \a in {30, 60}
    \draw[thick, dashed] (0, 0) -- (\a:1);
  \draw (30:1) node[anchor=west] {\tiny $T_1$};
  \draw (60:1) node[anchor=south west] {\tiny $T_2$};
  \draw (90:1.2) node[anchor=south] {\tiny $\lambda_2$};
  \draw (0:1.2) node[anchor=west] {\tiny $\lambda_1$};
  \foreach \a / \t in {
  15/{\tiny $\texttt{I}$},
  45/{\tiny $\texttt{II}$},
  75/{\tiny $\texttt{III}$}
  }
  \draw (\a:.8) node {\large \t};
  \foreach \a / \t in {
  135/{\tiny $\texttt{IV}$},
  225/{\tiny $\texttt{V}$},
  315/{\tiny $\texttt{VI}$}
  }
  \draw (\a:.3) node {\large \t};
\end{tikzpicture}
}
\subfigure[$\theta>0$, $\delta>0$, $\theta \delta < 1$]{
\begin{tikzpicture}[scale = 2]
  \draw[thick, dashed] (0, 0) -- (0:1.2);
  \draw[thick, dashed] (0, 0) -- (90:1.2);
  \draw[thick, dashed] (0, 0) -- (180:.5);
  \draw[thick, dashed] (0, 0) -- (270:.5);
  \foreach \a in {30, 60}
    \draw[thick, dashed] (0, 0) -- (\a:1);
  \draw (30:1) node[anchor=west] {\tiny $T_2$};
  \draw (60:1) node[anchor=south west] {\tiny $T_1$};
  \draw (90:1.2) node[anchor=south] {\tiny $\lambda_2$};
  \draw (0:1.2) node[anchor=west] {\tiny $\lambda_1$};
  \foreach \a / \t in {
  15/{\tiny $\texttt{I}$},
  45/{\tiny $\texttt{VII}$},
  75/{\tiny $\texttt{III}$}
  }
  \draw (\a:.8) node {\large \t};
  \foreach \a / \t in {
  135/{\tiny $\texttt{IV}$},
  225/{\tiny $\texttt{V}$},
  315/{\tiny $\texttt{VI}$}
  }
  \draw (\a:.3) node {\large \t};
\end{tikzpicture}
}
\subfigure[$\theta>0$, $\delta<0$]{
\begin{tikzpicture}[scale = 1.5 ]
  \draw[thick, dashed] (0, 0) -- (0:1.2);
  \draw[thick, dashed] (0, 0) -- (90:1.2);
  \draw[thick, dashed] (0, 0) -- (180:.5);
  \draw[thick, dashed] (0, 0) -- (270:.5);
  \foreach \a in {45, -45}
    \draw[thick, dashed] (0, 0) -- (\a:1);
  \draw (-45:1) node[anchor=west] {\tiny $T_2$};
  \draw (45:1) node[anchor=south west] {\tiny $T_1$};
  \draw (90:1.2) node[anchor=south] {\tiny $\lambda_2$};
  \draw (0:1.2) node[anchor=west] {\tiny $\lambda_1$};
  \foreach \a / \t in {
  15/{\tiny $\texttt{VII}$},
  65/{\tiny $\texttt{VIII}$},
  300/{\tiny $\texttt{VI}$},
  340/{\tiny $\texttt{IX}$}
  }
  \draw (\a:.8) node {\large \t};
  \foreach \a / \t in {
  135/{\tiny $\texttt{IV}$},
  225/{\tiny $\texttt{V}$}
  }
  \draw (\a:.3) node {\large \t};
\end{tikzpicture}
}
\caption{ The regions in the $\lambda_1$--$\lambda_2$
  plane with different dynamical behaviors. In region \texttt{V}, the basic
  steady state is locally asymptotically stable. In regions \texttt{IV} and
  \texttt{VI}, the system undergoes a supercritical Hopf bifurcation. The
  dynamics in regions \texttt{I}, \texttt{II} and \texttt{III} is the double
  Hopf bifurcation scenario and the details are given in
  \autoref{fig:double-hopf-fig2}. The lines $T_1$ and $T_2$ in the figure have
  slopes $1/\theta$ and $\delta$ (as defined in~\eqref{delta})
  respectively.\label{fig:double-hopf-diag}}%
\end{figure}
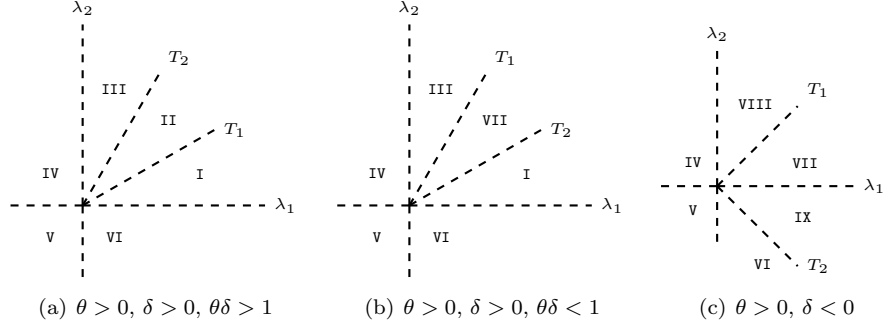

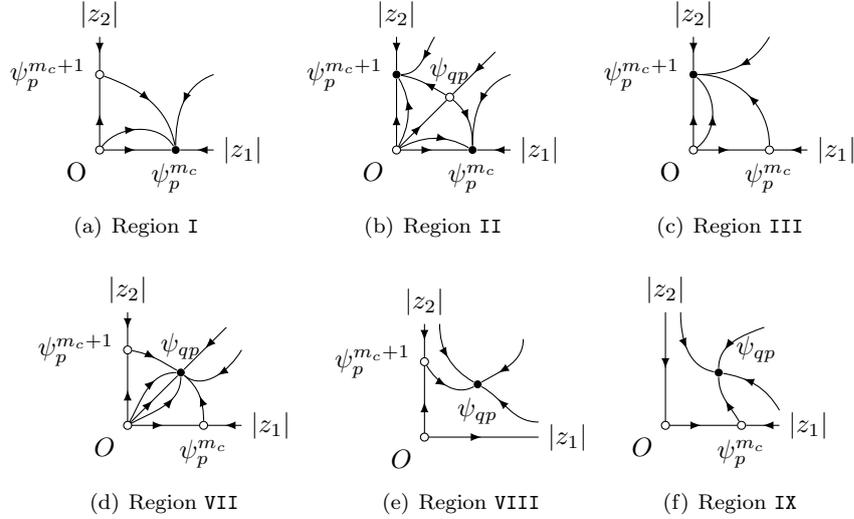
\begin{figure}[tb]
  \centering
  \tikzset{->-/.style={decoration={
        markings,
        mark=at position #1 with {\arrow[scale=1]{latex}}},postaction={decorate}}}
  \tikzset{-<-/.style={decoration={
        markings,
        mark=at position #1 with {\arrowreversed[scale=1]{latex}}},postaction={decorate}}}
  \subfigure[Region \texttt{I}]{
    \begin{tikzpicture}[scale=1]
      \node[circle,fill=black,inner sep=0pt,minimum size=3pt,label=below:{$\psi_{p}^{m_c}$}] (E1) at (1,0) {};
      \node[circle,fill=white, draw, inner sep=0pt,minimum size=3pt,label=left:{$\psi_{p}^{m_c+1}$}] (E2) at (0,1) {};
      \node[right] at (1.5,0) {$\abs{z_1}$};
      \node[above] at (0,1.5) {$\abs{z_2}$};
      \draw[->-=.5] (0,0) to (E2);
      \draw[->-=.5] (0,1.5) to (E2);
      \draw[->-=.5] (0,0) to (E1);
      \draw[->-=.5] (1.5,0) to (E1);
      \draw[->-=.5] (E2) to [out=340,in=100] (E1);
      \draw[->-=.5] (0,0) to [out=60,in=120] (E1);
      \draw[->-=.5] (1.5,1) to [out=200,in=90] (E1);
      \node[circle,fill=white, draw, inner sep=0pt,minimum size=3pt,label=below left:{O}] (E0) at (0,0) {};
    \end{tikzpicture}
    \label{fig:dm1gm1}
  }
  \subfigure[Region \texttt{II}]{
    \begin{tikzpicture}[scale=1]
      \node[circle,fill=black,inner sep=0pt,minimum size=3pt,label=below:{$\psi_{p}^{m_c}$}] (E1) at (1,0) {};
      \node[circle,fill=black,inner sep=0pt,minimum size=3pt,label=left:{$\psi_{p}^{m_c+1}$}] (E2) at (0,1) {};
      \node[circle,fill=white,draw, inner sep=0pt,minimum size=3pt,label=above:{$\psi_{qp}$}] (E3) at (.7,.7) {};
      \node[right] at (1.5,0) {$\abs{z_1}$};
      \node[above] at (0,1.5) {$\abs{z_2}$};
      \draw[->-=.5] (0,0) to (E2);
      \draw[->-=.5] (0,1.5) to (E2);
      \draw[->-=.5] (0,0) to (E1);
      \draw[->-=.5] (1.5,0) to (E1);
      \draw[->-=.5] (0,0) to (E3);
      \draw[->-=.5] (1.3,1.3) to (E3);
      \draw[->-=.5] (E3) to [out=150,in=-10] (E2);
      \draw[->-=.5] (E3) to [out=320,in=90] (E1);
      \draw[->-=.7] (0,0) to [out=30,in=150] (E1);
      \draw[->-=.7] (0,0) to [out=60,in=-60] (E2);
      \draw[->-=.5] (1.5,1) to [out=200,in=90] (E1);
      \draw[->-=.5] (.5,1.5) to [out=240,in=0] (E2);
      \node[circle,fill=white, draw, inner sep=0pt,minimum size=3pt,label=below left:{$O$}] (E0) at (0,0) {};
    \end{tikzpicture}
    \label{fig:dm1gm2}
  }
  \subfigure[Region \texttt{III}]{
    \begin{tikzpicture}[scale=1]
      \node[circle,fill=white, draw, inner sep=0pt,minimum size=3pt,label=below:{$\psi_{p}^{m_c}$}] (E1) at (1,0) {};
      \node[circle,fill=black,inner sep=0pt,minimum size=3pt,label=left:{$\psi_{p}^{m_c+1}$}] (E2) at (0,1) {};
      \node[right] at (1.5,0) {$\abs{z_1}$};
      \node[above] at (0,1.5) {$\abs{z_2}$};
      \draw[->-=.5] (0,0) to (E1);
      \draw[->-=.5] (0,1.5) to (E2);
      \draw[->-=.5] (0,0) to (E2);
      \draw[->-=.5] (1.5,0) to (E1);
      \draw[->-=.5] (E1) to [out=90,in=0] (E2);
      \draw[->-=.5] (0,0) to [out=30,in=320] (E2);
      \draw[->-=.5] (1,1.5) to [out=240,in=0] (E2);
      \node[circle,fill=white, draw, inner sep=0pt,minimum size=3pt,label=below left:{O}] (E0) at (0,0) {};
    \end{tikzpicture}
    \label{fig:dm1gm3}
  }

  \subfigure[Region \texttt{VII}]{
    \begin{tikzpicture}[scale=1]
      \node[circle,fill=white, draw, inner sep=0pt,minimum size=3pt,label=below:{$\psi_{p}^{m_c}$}] (E1) at (1,0) {};
      \node[circle,fill=white, draw, inner sep=0pt,minimum size=3pt,label=left:{$\psi_{p}^{m_c+1}$}] (E2) at (0,1) {};
      \node[circle,fill=black, inner sep=0pt,minimum size=3pt,label=above:{$\psi_{qp}$}] (E3) at (.7,.7) {};
      \node[right] at (1.5,0) {$\abs{z_1}$};
      \node[above] at (0,1.5) {$\abs{z_2}$};
      \draw[->-=.5] (0,0) to (E2);
      \draw[->-=.5] (0,1.5) to (E2);
      \draw[->-=.5] (0,0) to (E1);
      \draw[->-=.5] (1.5,0) to (E1);
      \draw[->-=.5] (0,0) to (E3);
      \draw[->-=.5] (1.3,1.3) to (E3);
      \draw[-<-=.5] (E3) to [out=150,in=-10] (E2);
      \draw[-<-=.5] (E3) to [out=320,in=90] (E1);
      \draw[->-=.7] (0,0) to [out=30,in=270] (E3);
      \draw[->-=.7] (0,0) to [out=60,in=180] (E3);
      \draw[->-=.5] (1.5,1) to [out=240,in=320] (E3);
      \node[circle,fill=white, draw, inner sep=0pt,minimum size=3pt,label=below left:{$O$}] (E0) at (0,0) {};
    \end{tikzpicture}
    \label{fig:dm1gm4}
  }
  \subfigure[Region \texttt{VIII}]{
    \begin{tikzpicture}[scale=1]
      \node[circle,fill=white, draw, inner sep=0pt,minimum size=3pt,label=left:{$\psi_{p}^{m_c+1}$}] (E2) at (0,1) {};
      \node[circle,fill=black, inner sep=0pt,minimum size=3pt,label=below:{$\psi_{qp}$}] (E3) at (.7,.7) {};
      \node[right] at (1.5,0) {$\abs{z_1}$};
      \node[above] at (0,1.5) {$\abs{z_2}$};
      \draw[->-=.5] (0,0) to (E2);
      \draw[->-=.5] (0,1.5) to (E2);
      \draw[->-=.5] (0,0) to (1.5, 0);
      \draw[->-=.5] (E2) to [out=-60,in=210] (E3);
      \draw[->-=.5] (.2, 1.5) to [out=270,in=150] (E3);
      \draw[->-=.7] (1.3, 1.3) to [out=270,in=30] (E3);
      \draw[->-=.7] (1.5,.2) to [out=180,in=330] (E3);
      \node[circle,fill=white, draw, inner sep=0pt,minimum size=3pt,label=below left:{$O$}] (E0) at (0,0) {};
    \end{tikzpicture}
    \label{fig:dm1gm5}
  }
  \subfigure[Region \texttt{IX}]{
    \begin{tikzpicture}[scale=1]
      \node[circle,fill=white, draw, inner sep=0pt,minimum size=3pt,label=below:{$\psi_{p}^{m_c}$}] (E1) at (1,0) {};
      \node[circle,fill=black, inner sep=0pt,minimum size=3pt, label={[label distance=.05cm]10:{$\psi_{qp}$}}] (E3) at (.7,.7) {};
      \node[right] at (1.5,0) {$\abs{z_1}$};
      \node[above] at (0,1.5) {$\abs{z_2}$};
      \draw[->-=.5] (0,0) to (E1);
      \draw[->-=.5] (1.5,0) to (E1);
      \draw[-<-=.5] (0,0) to (0, 1.5);
      \draw[->-=.5] (E1) to [out=120,in=270] (E3);
      \draw[->-=.7] (1.3, 1.3) to [out=200,in=90] (E3);
      \draw[->-=.5] (.2, 1.5) to [out=270,in=170] (E3);
      \draw[->-=.7] (1.5,.2) to [out=120,in=350] (E3);
      \node[circle,fill=white, draw, inner sep=0pt,minimum size=3pt, label=below left:{$O$}] (E0) at (0,0) {};
    \end{tikzpicture}
    \label{fig:dm1gm6}
  }

  \caption{The dynamics in the regions given in the first quadrant of
    \autoref{fig:double-hopf-diag}. $\psi_{p}^{m_c}$, $\psi_{p}^{m_c+1}$ are time-periodic with zonal wavenumbers $m_c$ and $m_c+1$ respectively. $\psi_{qp}$ is the quasiperiodic solution given in~\eqref{per-qper}.}
  \label{fig:double-hopf-fig2}
\end{figure}

\begin{theorem}\label{thm2}
  Assume that the conditions of \autoref{lemma2} as well as the condition~\eqref{theta-delta-condition} hold.
  Then the equations \eqref{eq:main-abstract-form} undergo a continuous transition at $\Psi=\Psi_c$, and an $S^{3}$ attractor $\Sigma$ bifurcates on $\Psi>\Psi_c$, which converges to $\mathbf{0}$ as $\Psi$ approaches to $\Psi_c$ from right. Depending on the values of $\theta$ and $\delta$, there are three transition scenarios as shown in \autoref{fig:double-hopf-diag}.
  In each scenario, near the onset of transition $(\lambda_1, \lambda_2) = (0, 0)$, the $\lambda_1-\lambda_2$ plane is dissected into several regions with distinct topological structures for the attractor $\Sigma$ as given in~\autoref{fig:double-hopf-fig2}.
\end{theorem}

\begin{figure}[t]
  \centering
  \includegraphics[height=0.5\textwidth, width=.7\textwidth]{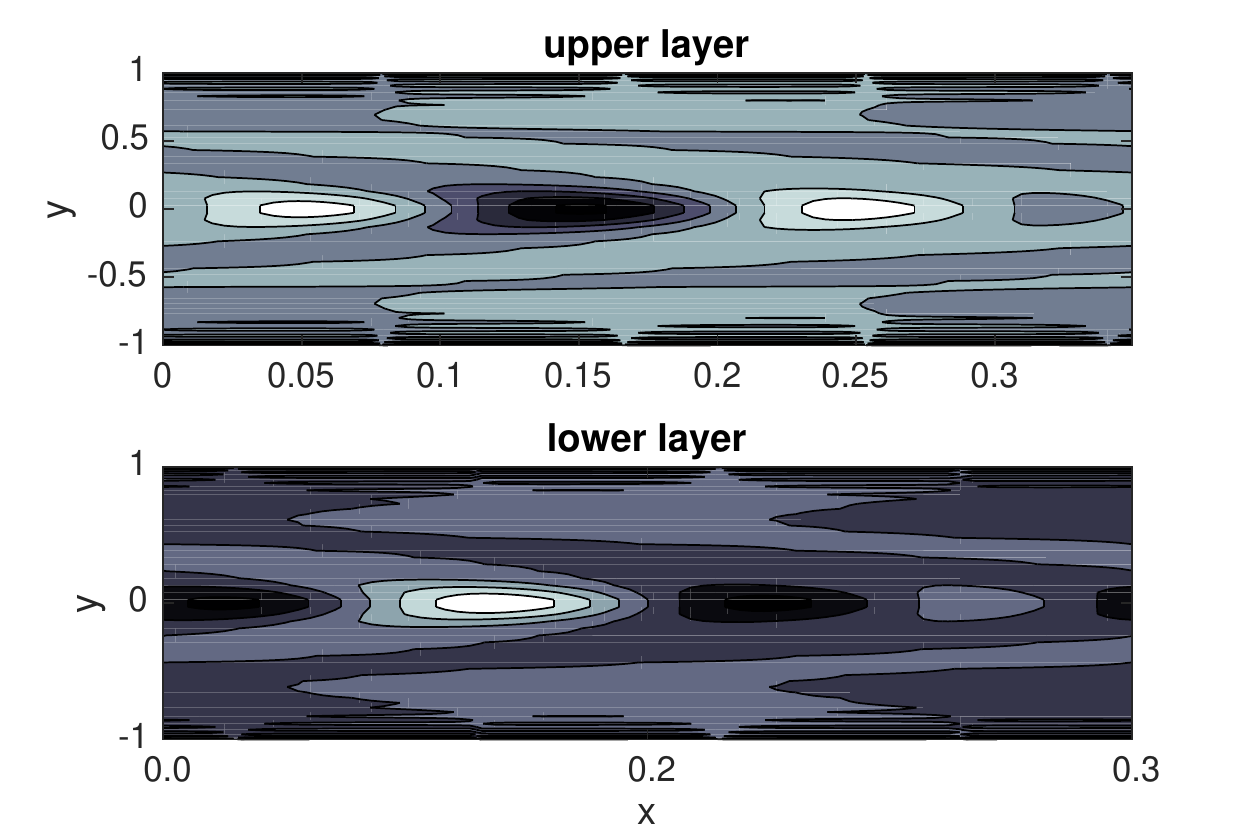}
  \caption{Real part of the upper and lower layers of the time quasi-periodic solution $\psi_{qp}$ given in~\autoref{thm2} at $t=0$ for $k=1$ and $a = 5.722$ where the first two critical wavenumbers are $m_c=1$ and $m_c=2$.}
\label{fig:dhk1}
\end{figure}

\begin{remark}%
  \begin{enumerate}
    \item If $z_2=0$ or $z_1 = 0$, the equations~\eqref{req-m2} reduce to the
      equation~\eqref{req-m1} with $A=P$ or $D=P$ respectively. Thus
      \autoref{lemma1} and \autoref{thm1} are special cases of
      \autoref{lemma2} and \autoref{thm2} respectively.
    \item We note that the features of the spatial structures of upper vs lower
      layer of the bifurcated periodic solutions in~\autoref{fig:crit_evec} and
      the quasi-periodic solution in~\autoref{fig:dhk1} do not alter much. We
      expect that the situation would be different if bottom topography is included.
  \end{enumerate}%
\end{remark}

The transition scenario of double Hopf transition is given by~\autoref{thm2}
by~\autoref{fig:double-hopf-diag} and~\autoref{fig:double-hopf-fig2}. We find
that near the onset of transition, depending on the fluctuations the basic state {experiences a
transition} either to a time periodic solution or {to a} quasi periodic solution.
Our results in \autoref{tab:doublehopf} show that {the three scenarios sketched} in \autoref{fig:double-hopf-diag} are realizable.

In particular, near a double Hopf transition point, one of the following three possibilities must occur post transition, $\Psi > \Psi_c$:
\begin{itemize}
  \item[(i)] there is only a single stable limit cycle,
  \item[(ii)] there are two distinct stable limit cycles, and an unstable quasi-periodic solution
  \item[(iii)] there is a stable quasi-periodic solution and either one or two unstable limit cycles.
\end{itemize}
For the double Hopf transition, \autoref{thm2} basically tells that all of the above local structures, the time-periodic solution and the 2D torus, if they exist, reside in a local attractor homeomorphic to the three dimensional sphere. The existence of this attracting 3D sphere is guaranteed by the attractor bifurcation theorem; see {\cite[Theorem 6.1]{ma2005}}.

\section{Summary and Discussion}\label{Sec_discuss}

In this paper, we investigated the stability of a parallel zonal jet  forced  by a Kolmogorov-type
wind stress in a periodic zonal channel, using  a two-layer quasi-geostrophic (QG) model.
This  problem is, in terms  of complexity, situated between the horizontally unbounded problem
\cite{Phillips1951,VanderVaart1997} and the fully bounded gyre  problem \cite{Dijkstra1997}.
More precisely, the effect of boundaries is captured by the interactions of only a few modes
(solutions of the  linear stability problem), while still keeping a parallel flow which is  connected
to the wind-stress field.

Our results show,  {in the continuity of earlier works} \cite{kieu2018,cai2017,dijkstra2015,lu2020}, {that} the zonal shear flow is linearly stable if its maximal amplitude
 $\Psi$, or equivalently the maximal amplitude $\tau$ of the wind-stress curl {(see \eqref{eq:1})},  is below
 a critical threshold $\Psi_c$. Moreover, as this critical threshold is exceeded, generically
 a Hopf bifurcation occurs; {see \autoref{thm1}}. {To derive this result, we approached} the problem {by means of} dynamic transition theory determining
all the attractors near {the onset of instability. These attractors then describe how the onset of linear instability translates into the emergence of nonlinear patterns for the actual flow; see \eqref{thm1 solution}.}

 
 {
 {Our approach allows for computing the coefficients of the Hopf reduced equation \eqref{req-m1} by means of explicit formulas involving the interactions between the critical mode and the (zonal) stable ones through the nonlinear terms; see
 \eqref{Eq_P}-\eqref{Eq_P2}. 
 A numerical examination of these coefficients reveals} that instead of infinitely many {(stable)} modes
 which {would affect} the type of transition (supercritical vs subcritical Hopf), the transition is {in fact} determined
by the interactions {between the conjugate pair of critical modes losing stability and} only the first few zonally homogeneous
($m=0$) modes. {That the interactions with the zonal-mean modes are {the determining factors governing the flow patterns emerging after the onset of instability may be viewed as consistent with other works that have shown (in more turbulent regimes)} that 
{baroclinic turbulence statistics can be well recovered for certain geophysical flows by} omitting the {effects of the} interactions {among the higher-order modes (eddy-eddy interactions); see e.g.~\cite{panetta1988baroclinic,gorman2007,srinivasan2012zonostrophic,chemke2016effect}.} }
}

We also investigated the double Hopf bifurcation scenario which takes place at critical length
scales where four modes with consecutive wavenumbers become critical. By a rigorous center
manifold analysis, we obtain the coefficients of the 4D-ODE system. Our results show that for
the parameters we have considered, there exists a quasi-periodic solution {that bifurcates. This quasi-periodic solution} is a linear
combination of two periodic solutions and may be stable depending on the parameters; {see \autoref{Sec_doubleHopf}}. From
a transition point of view, in the double Hopf transition, an attractor homeomorphic to 3D sphere
bifurcates; {see \autoref{thm2}}. This attractor contains stable/unstable limit cycles and an stable/unstable
invariant torus ({supporting a} quasi-periodic solution).

{Overall, our} results add more details to the nonlinear development of baroclinic instabilities on a
non-constant parallel zonal jet, in that the periodic {solutions} can become unstable to torus
bifurcations and give rise to quasi-periodic behavior. Such a scenario was also found
for the barotropic double gyre flow \cite{VanderVaart2002}, but only in a weakly
nonlinear framework using a set of (reduced) amplitude equations. The transition scenario
found for the zonally periodic zonal jet   is likely to be more  relevant for the ocean circulation
than the sideband  instabilities in the zonally unbounded  zonal jet case which require a
nearby band of  wavenumbers to be unstable.

In our set-up, the linear friction coefficient in the upper layer is relatively large and as
explained needed to balance the vorticity input by the wind stress for generating the
zonal jet. When this friction is decreased, more modes will become unstable near the
critical point and their interactions {are expected} to give a detailed view on the {development
of the unstable modes into ocean vortices, or eddies, due to baroclinic instabilities.} {In that respect,} the dynamic transition theory {adopted here}, {along} with extensions {based on the variational approach of \cite{CLM16_Lorenz9D,Chekroun2019} and its stochastic rectification \cite{Chekroun2021c}  to handle regimes beyond the onset of instability}, provides a way forward to develop a mathematical theory of such ocean-eddy  formation processes. {The results derived here combined with recent advances in the rigorous analysis of transitions arising in stochastically driven flows \cite{chekroun2022transitions} open new prospects  for the study of regime transitions in stochastically forced ocean models \cite{SFL01,SD13}.}

\appendix

\renewcommand{\thefigure}{A\arabic{figure}}
\setcounter{figure}{0}

\section{Proof of~\autoref{lemma1} and~\autoref{thm1}}\label{Sec_AppB}
We first proceed with the proof of \autoref{lemma1} For this, we denote the
adjoint modes by
\begin{displaymath}
  \psi^{*}_{m, j} = e^{i \alpha_{m} x} Y_{m}^{*}(y).
\end{displaymath}
We denote the critical eigenmode and the critical eigenvalue by
\begin{displaymath}
  \psi_c = \psi_{m_c, 1}, \qquad
  \sigma_c = \sigma_{m_c, 1}.
\end{displaymath}
We denote the bilinear operator $\mathcal{G}$ as
\begin{displaymath}
  \mathcal{G}(u) = G_{2}(u, u)
\end{displaymath}
where $G_{2}(u, v)$ is linear in each component. Let us define now
\begin{equation}
  \label{Gs}
  G_{s}(u, v) = G_{2}(u, v) + G_{2}(v, u).
\end{equation}
The center part of the solution is
\begin{equation}
  \label{centerpart}
  u_{c} = z(t) \psi_{c} + c.c.
\end{equation}
where c.c. stands for complex conjugate of the terms before.

The evolution of $z(t)$ near the onset of transition is obtained by the
projection onto the critical mode $\psi_c$.
\begin{equation}
  \label{reduced1}
  \dot{z} = \sigma_{c} z + \frac{ 1 }{ \langle \mathcal{M} \psi_{c}, \psi_{c}^{*} \rangle }
  \langle \mathcal{G}( u_{c} + \Phi ), \psi_{c}^{*} \rangle.
\end{equation}
where $\Phi$ is the center manifold function. We will obtain its quadratic
approximation $\Phi_2$ given by
\begin{displaymath}
  \Phi = \Phi_2(z, \overline{z}) + o(2) \\
\end{displaymath}
Here
\begin{displaymath}
  o(n) = o \left( \abs{ (z, \overline{z}) }^{n} \right)
\end{displaymath}
denotes higher than $n$-th order terms in $z$, $\overline{z}$.

Using the notation \eqref{Gs}, the reduced equation \eqref{reduced1} can be
written
\begin{equation}
  \label{reduced3}
  \dot{z} = \sigma_{c} z + \frac{ 1 }{ \langle \mathcal{M} \psi_{c}, \psi_{c}^{*} \rangle } \langle G_{s}( u_{c}, \Phi_2 ), \psi_{c}^{*} \rangle + o(3).
\end{equation}

To obtain a closed system, we need to approximate the center manifold function.
The approximation of the center manifold in this case reads,
see~\cite{sengul2018},
\begin{equation}
  \label{CM1}
  \begin{aligned}
    & \Phi_{2} = (2 \sigma_{c} - \mathcal{L})^{-1} \Pi_{s} G_2(z \psi_c, z \psi_c) +
    (\sigma_c + \overline{ \sigma }_c - \mathcal{L})^{-1} \Pi_{s} G_{2}(z \psi_{c}, \overline{z \psi_{c}}) + c.c.
  \end{aligned}
\end{equation}
where $\mathcal{L} = \Pi_{s} \mathcal{M}^{-1} \mathcal{N}$ and $\Pi_{s}$ is the projection
on the stable space. Using the formula~\eqref{CM1}, we obtain the following expansion of
the center manifold {(see also \cite[Theorem 2]{Chekroun2019})}
\begin{equation}
  \label{CM2}
  \Phi_{2} =
  z^{2} \sum_{j \ge 1} g_{2m_{c}, j}  \psi_{2m_{c}, j} +
  \abs{ z }^{2} \sum_{j \ge 1} g_{0, j}  \psi_{0, j} + c.c.
\end{equation}
Here
\begin{equation}
  \label{g}
  \begin{aligned}
    & g_{0, j} = \frac{ 1 }{ (\sigma_c + \overline{ \sigma }_c -\sigma_{0, j})  \langle \mathcal{M} \psi_{0,j}, \psi_{0,j}^{*} \rangle }
    \langle G_{2}(\psi_{c}, \overline{\psi_{c}}), \psi_{0, j}^{*}  \rangle \\
    & g_{2m_{c}, j} = \frac{ 1 }{ ( 2\sigma_{c} - \sigma_{2 m_{c}, j} ) \langle \mathcal{M} \psi_{2m_{c}, j}, \psi_{2m_{c},j}^{*} \rangle } \langle G_{2}(\psi_{c}, \psi_{c}), \psi_{2m_{c}, j}^{*} \rangle,
  \end{aligned}
\end{equation}
are the coefficients of the center manifold function.

We write~\eqref{reduced3} as~\eqref{req-m1}, that is
\begin{displaymath}
  \dot{z} = \sigma_c z + P z \abs{ z }^2 + o(3).
\end{displaymath}
which finishes the proof of~\autoref{lemma1}.

Recalling the definition of $G_s$ given in \eqref{Gs},  the transition
number $P$ can then be written as
\begin{equation}\label{Eq_P}
  P = P_0 + P_2,
\end{equation}
where
\begin{equation}\label{Eq_P0}
  P_{0} = \sum_{j \ge 1} P_{0, j}, \qquad
  P_{0, j} = \frac{ 1 }{ \langle \mathcal{M} \psi_{c}, \psi_{c}^{*} \rangle } ( g_{0, j} + c.c.)
  \langle G_{s}(\psi_{c}, \psi_{0, j}), \psi_{c}^{*} \rangle,
\end{equation}
denotes the contribution of {the zero-wavenumber (stable)} modes $\psi_{0, j}$ while
\begin{equation}\label{Eq_P2}
  P_{2} = \sum_{j \ge 1} P_{2, j}, \qquad
  P_{2, j} = \frac{ 1 }{ \langle \mathcal{M} \psi_{c}, \psi_{c}^{*} \rangle } g_{2m_{c}, j}
  \langle G_{s}(\overline{\psi_{c}}, \psi_{2m_{c}, j}), \psi_{c}^{*} \rangle,
\end{equation}
denotes the contribution of {the} modes $\psi_{2m_{c}, j}$ on the transition
number respectively. The {transition type} depends on the real part of the transition
number $P$. The proof of~\autoref{thm1}
follows from the standard Hopf bifurcation analysis of the reduced equation.

\section{Proof of~\autoref{lemma2} and~\autoref{thm2}}\label{Sec_AppD}
As the reduction in the case of~\eqref{PES2} is similar to the case
of~\eqref{PES} given in the previous section, we will only mention the
differences between these two cases. Under the assumption~\eqref{PES2}, we write
the center part of the solution as
\begin{displaymath}
  u_c = z_1(t) \psi_1 + z_2(t) \psi_2 + c.c.
\end{displaymath}
where the first two critical modes are
\begin{displaymath}
  \psi_1 = \psi_{m_c, 1}, \quad \psi_2 = \psi_{m_c+1, 1}, \quad
  \psi_{-1} = \psi_{-m_c, 1}, \quad \psi_{-2} = \psi_{-m_c-1, 1}.
\end{displaymath}
with corresponding eigenvalues
\begin{displaymath}
  \sigma_1 = \sigma_{m_c, 1}, \quad   \sigma_2 = \sigma_{m_c+1, 1}, \quad
  \sigma_{-1} = \sigma_{-m_c, 1}, \quad   \sigma_{-2} = \sigma_{-m_c-1, 1}
\end{displaymath}
The equation~\eqref{reduced3} becomes the system
\begin{equation}\label{red-dh1}
  \dot{z}_j = \sigma_j z_j + \frac{ 1 }{ \langle \mathcal{M} \psi_j, \psi_j^* \rangle }
  \langle \mathcal{G}(u_c + \Phi_2), \psi_j^* \rangle + o(3), \qquad j = 1,2.
\end{equation}
and the center manifold function~\eqref{CM1} is replaced by
\begin{displaymath}
  \Phi_2 = \sum_{\abs{ j }, \abs{ k }=1}^{2} z_j z_k \Psi_{j, k} + o(2), \qquad
  \Psi_{j, k} = (\sigma_j + \sigma_k - \mathcal{L})^{-1} \Pi_s G_2(\psi_j, \psi_k),
\end{displaymath}
{where $\Pi_s$ denotes the projector onto the stable subspace.}

Now, the equations \eqref{red-dh1} become \eqref{req-m2} with the coefficients
defined below:
\begin{equation}\label{ABCD}
  \begin{aligned}
    & A = g_{1, 1, -1, 1} + g_{1, -1, 1, 1} + g_{-1, 1, 1, 1} \\
    & B = g_{1, 2, -2, 1} + g_{1, -2, 2, 1} + g_{2, 1, -2, 1} + g_{2, -2, 1, 1} + g_{-2, 1, 2, 1} + g_{-2, 2, 1, 1}\\
    & C = g_{2, 1, -1, 2} + g_{2, -1, 1, 2} + g_{1, 2, -1, 2} + g_{1, -1, 2, 2} + g_{-1, 2, 1, 2} + g_{-1, 1, 2, 2}\\
    & D = g_{2, 2, -2, 2} + g_{2,-2, 2, 2} + g_{-2, 2, 2 ,2}\\
    & g_{i, j, k, l} = \frac{ 1 }{ \langle \mathcal{M} \psi_l, \psi_l^* \rangle }
    \langle G_s( \psi_i, \Psi_{j, k}, \psi_l^* ) \rangle, \\
    & \Psi_{j ,k} = (\sigma_j + \sigma_k - \mathcal{L})^{-1}
    \Pi_s G_2(\psi_j,  \psi_k)
  \end{aligned}
\end{equation}
We note that the above coefficients contain only $g_{i, j, k, l}$ for which
$i+j+k = l$. The expansion of the center manifold coefficients can be written
more explicitly as {(see also \cite[Theorem 2]{Chekroun2019})}
\begin{displaymath}
  \Psi_{j ,k} = \sum_{i=1}^{\infty}
  \frac{ \langle  G_2(\psi_{m_j}, \psi_{m_k}) , \psi^*_{m_j+m_k, i} \rangle  }
  { \langle \mathcal{M} \psi_{m_j+m_k, i}, \psi^*_{m_j+m_k, i} \rangle }
  (\sigma_j + \sigma_k - \sigma_{m_j + m_k})^{-1} \psi_{m_j+m_k, i}
\end{displaymath}

Now we analyze the equations~\eqref{req-m2} by first putting them in polar form
\begin{displaymath}
  z_j = \rho_i e^{i \gamma_j}, \quad j = 1, 2
\end{displaymath}
which yields
\begin{equation}\label{dh-polar}
  \begin{aligned}
    & \dot{ \rho_1 } = Re(\sigma_1) \rho_1 + \rho_1 (Re(A) \rho_1^2 + Re(B) \rho_2^2) + \text{ h.o.t. }\\
    & \dot{ \rho_2 } = Re(\sigma_2) \rho_2 + \rho_2 (Re(C) \rho_1^2 + Re(D) \rho_2^2) + \text{ h.o.t. }
  \end{aligned}
\end{equation}
and
\begin{displaymath}
  \begin{aligned}
    & \dot{ \gamma_1 } = Im(\sigma_1) + \text{ h.o.t. } \\
    & \dot{ \gamma_2 } = Im(\sigma_2) + \text{ h.o.t. }
  \end{aligned}
\end{displaymath}
For the specific case of~\eqref{theta-delta-condition}, the
equations~\eqref{dh-polar} always admit the solutions which represent the
periodic solutions
\begin{align*}
  & (\rho_1, \rho_2) = \left(-\frac{ Re(\sigma_{1}) }{ Re(A) }, 0 \right), \\
  & (\rho_1, \rho_2) = \left(0, -\frac{ Re(\sigma_{2}) }{ Re(D) } \right),
\end{align*}
with respective eigenvalues
\begin{align*}
  & \kappa_1 = -2 \sigma_1, \qquad \kappa_2 = \sigma_2 - \delta \sigma_1 \\
  & \kappa_1 = -2 \sigma_2, \qquad \kappa_2 = \sigma_1 - \theta \sigma_2 \\
\end{align*}
Also, the equations~\eqref{dh-polar} admit the following solution which
represents a quasi-periodic solution
\begin{displaymath}
  (\rho_1, \rho_2) = \left( \frac{ \sigma_{1} - \theta \sigma_{2} }{ Re(A) (\theta \delta - 1) }, \frac{ \sigma_{2} - \delta \sigma_{1} }{ Re(D) (\theta \delta - 1) } \right).
\end{displaymath}
Since the Jacobian matrix of the right hand side of~\eqref{dh-polar} at the quasi-periodic solution has determinant
\begin{displaymath}
  \frac{ -4 ( \sigma_2 - \delta \sigma_1 ) ( \sigma_1 - \theta \sigma_2 )  }{ \theta \delta - 1 }.
\end{displaymath}
With this information, the transition scenarios summarized in \autoref{fig:double-hopf-diag} and \autoref{fig:double-hopf-fig2} can be obtained by a standard analysis.
To prove the claim on the bifurcation of an $S^3$-attractor, we need to prove that $(\rho_1, \rho_2) = (0, 0)$ is locally stable equilibrium of~\eqref{dh-polar} at $\Psi = \Psi_c$, that is when $\Re(\sigma_1) = \Re (\sigma_2) = 0$.
In this case by assumption~\eqref{theta-delta-condition}, from \eqref{dh-polar}, we can obtain
\begin{displaymath}
  \frac{ d }{ dt } (\rho_1^2 + \rho_2^2) = Re(A) \rho_1^4 + (\Re(B) + \Re(C))\rho_1^2 \rho_2^2 + \Re(D) \rho_2^4 < 0
\end{displaymath}
which proves the claim.

\section{Numerical treatment of the linear stability problem}\label{Sec_AppA}

To solve the eigenvalue problem numerically, we first plugin the ansatz
\begin{equation}
  \label{eq:ansatz}
  \psi(x, y) = e^{i \alpha_{m} x} Y_{j}(y), \qquad j \in \mathbb{N}, m \in \mathbb{Z}, \quad
  \alpha_{m} := a m \pi.
\end{equation}
into the eigenvalue problem
\begin{displaymath}
  \sigma \mathcal{M} \psi(x,y) = \mathcal{N}(y) \psi(x,y),
\end{displaymath}
to obtain
\begin{equation}
  \label{eq:eigenvalue problem}
  \sigma \widetilde{\mathcal{M}} Y(y) = \widetilde{\mathcal{N}}(y) Y(y), \qquad Y(y) = (Y_{1}(y), Y_{2}(y))
\end{equation}
Here the linear operators $\widetilde{\mathcal{M}}$ and
$\widetilde{\mathcal{N}}(y)$ are defined as
\begin{equation}
  \label{eq:M matrix}
  \widetilde{\mathcal{M}} =
  \begin{bmatrix}
    \Delta_{m} - F_{1} & F_{1} \\
    F_{2} & \Delta_{m} - F_{2}
  \end{bmatrix}, \qquad
  \widetilde{\mathcal{N}}(y) =
  \begin{bmatrix}
    N_{11} & N_{12} \\ N_{21} & N_{22}
  \end{bmatrix}
\end{equation}
where
\begin{equation}
  \label{eq:7}
  \begin{aligned}
    & N_{11} = c_{1} \cos k \pi y \left( (k \pi)^{2} + \Delta_{m} \right) - \beta i \alpha_{m} - r_{1} \Delta_{m} \\
    & N_{12} = c_{1} F_{1} \cos k \pi y \\
    & N_{21} = 0 \\
    & N_{22} = -c_{1} F_{2} \cos k \pi y - \beta i \alpha_{m} - r_{2} \Delta_{m}
  \end{aligned}
\end{equation}
\begin{displaymath}
  \Delta_{m} = D^{2} - \alpha_{m}^{2}, \qquad D = \frac{\partial }{\partial y}
\end{displaymath}
\begin{displaymath}
  c_{1} = \Psi k \pi i \alpha_{m}.
\end{displaymath}

The eigenvalue problem~\eqref{eq:eigenvalue problem} is supplemented with the
following boundary conditions
\begin{equation}
  \label{eq:eigenvalue problem BC}
  Y_{i}(\pm 1) = D^{2}Y_{i} (\pm 1) = 0, \qquad i=1,2.
\end{equation}

We use Legendre-Galerkin method to discretize and solve the~\eqref{eq:eigenvalue
  problem} with boundary conditions~\eqref{eq:eigenvalue problem BC}. We refer
to~\cite{shen2011} for the details of the Legendre-Galerkin method and
to~\cite{dijkstra2015} for its use in dynamical transition problems.

Let $\{ L_{j} \}$ be the Legendre polynomials and consider compact combinations
of the Legendre polynomials
\begin{displaymath}
  f_{j}(y) = L_{j}(y) + \sum_{k=1}^{4} c_{jk} L_{j+k}(y)
\end{displaymath}
with $c_{jk}$ chosen so that $f_{j}$ satisfy the boundary
conditions~\eqref{eq:eigenvalue problem BC}, i.e.
\begin{displaymath}
  f_{j}(\pm 1) = D^{2}f_{j}(\pm 1) = 0.
\end{displaymath}

To discretize the eigenvalue problem, we plug
\begin{equation} \label{Ny}
  Y_{i}^{N_y}(y) = \sum_{j=0}^{N_y-1} y_{j}^{(i)} f_{j}(y), \qquad
  \widehat{Y}_{i} = [ y_{0}^{(i)}, \dots, y_{N_y-1}^{(i)} ]^{T}, \qquad i = 1, 2.
\end{equation}
into \eqref{eq:eigenvalue problem} to obtain
\begin{equation}
  \label{eq:dep}
  \sigma
  \begin{bmatrix}
    \widehat{\Delta}_{m} - F_{1} A_{3} & F_{1} A_{3} \\
    F_{2} A_{3} & \widehat{\Delta}_{m} - F_{2} A_{3}
  \end{bmatrix}
  \begin{bmatrix}
    \widehat{Y}_{1} \\ \widehat{Y}_{2}
  \end{bmatrix} =
  \begin{bmatrix}
    \widehat{N}_{11} & \widehat{N}_{12} \\ \widehat{N}_{21} & \widehat{N}_{22}
  \end{bmatrix}
  \begin{bmatrix}
    \widehat{Y}_{1} \\ \widehat{Y}_{2}
  \end{bmatrix}
\end{equation}

\begin{equation}
  \label{eq:6}
  \begin{aligned}
    & \widehat{N}_{11}= c_{1}(k \pi)^{2} A_{5}
    + c_{1}\left(A_{4}^{T}-\alpha_{m}^{2} A_{5}\right)
    - \beta i \alpha_{m} A_{3}
    - r_{1} \widehat{\Delta}_{m} \\
    & \widehat{N}_{12} = c_{1} F_{1} A_{5} \\
    & \widehat{N}_{21} = 0 \\
    & \widehat{N}_{22} = -c_{1} F_{2} A_{5}-\beta i \alpha_{m} A_{3}
    - r_2 \widehat{\Delta}_{m}
  \end{aligned}
\end{equation}

Here
\begin{equation}
  \label{eq:2}
  \begin{aligned}
    & A_{1} = (D^{4} f_{j}, f_{k}), \qquad A_{2} = (D^{2} f_{j}, f_{k}), \qquad A_{3} = (f_{j}, f_{k}), \\
    & A_{4} = (\cos k\pi y D^{2}f_{j}, f_{k}), \qquad A_{5} = (\cos k \pi y f_{j}, f_{k}), \\
    & \widehat{\Delta}_{m} = A_{2}-\alpha_{m}^{2} A_{3},
  \end{aligned}
\end{equation}
with { $(f, g) = \int_{-1}^{1} f (y) g (y) \d y$}. The explicit expression of the matrices
$A_{i}$, $i=1,\dots,5$ can be found in \cite{dijkstra2015}.

\section{{ Practical aspects for the calculation of} the transition number}\label{Sec_AppC}
The practical calculation of the $P_0$-term in \eqref{Eq_P0} and the $P_2$-term in \eqref{Eq_P2}, boils down to the efficient calculation of the inner and trilinear products involved therein. In that respect, we provide here explicit expressions of the latter. They are given by
\begin{displaymath}
  \begin{aligned}
    \langle \mathcal{M} \psi_{m, j}, \psi_{m, j}^{*} \rangle = i \alpha_{m} & \int_{-1}^{1}
    \left( (D^{2} - \alpha_{m}^{2} - F_{1}) Y_{m, j}^{1} + F_{1} D Y_{m, j}^{2} \right) \overline{ Y_{m, j}^{* 1}} \d y \\
    + i \alpha_{m} & \int_{-1}^{1}
    \left( (D^{2} - \alpha_{m}^{2} - F_{2}) Y_{m, j}^{2} + F_{2} D Y_{m, j}^{1} \right) \overline{ Y_{m, j}^{* 2}} \d y,
  \end{aligned}
\end{displaymath}
and
\begin{displaymath}
  \langle G_{2} (\psi_{m,j}, \psi_{n, k}), \psi_{p, l}^{*} \rangle =
  -\delta_{m+n-p} \left( \int_{-1}^{1} G_{2}^{1} \overline{ Y_{p,l}^{*, 1} } + G_{2}^{2} \overline{ Y_{p,l}^{*, 2} } \right) \d y,
\end{displaymath}

\begin{displaymath}
  \begin{aligned}
    G_{2}^{1} =  i \alpha_{m} & Y_{m, j}^{1} \left( D (D^{2} - \alpha_{n}^{2} - F_{1}) Y_{n, k}^{1} + F_{1} D Y_{n,k}^{2} \right) \\
    &- i \alpha_{n} \left( (D^{2} - \alpha_{n}^{2} - F_{1}) Y_{n, k}^{1} + F_{1} Y_{n,k}^{2} \right) D Y_{m, j}^{1} \\
    G_{2}^{2} = i \alpha_{m} & Y_{m, j}^{2} \left( D (D^{2} - \alpha_{n}^{2} - F_{2}) Y_{n, k}^{2} + F_{2} D Y_{n,k}^{1} \right) \\
    & - i \alpha_{n} \left( (D^{2} - \alpha_{n}^{2} - F_{2}) Y_{n, k}^{2} + F_{2} Y_{n,k}^{1} \right) D Y_{m, j}^{2}.
  \end{aligned}
\end{displaymath}

{ In practice, the integrals can be evaluated by any commonly used quadrature rules in which the values of the integrand are evaluated at quadrature points. In our calculations, we use
\begin{displaymath}
  \int_{-1}^{1} f(y) \d y = \sum_{n=0}^{N_y} f(y_{n}) \omega_{n},
\end{displaymath}
where $y_{n}$ and $\omega_{n}$ are Legendre-Gauss-Lobatto quadrature points and
weights respectively.}

\section{Model Parameters}\label{Sec_AppE}

\begin{table}[H]
  \centering
  \begin{tabular}{|l|l|l|}
    \hline
    $L_y$ & meridional length scale & $10^6$ m  \\
    $H_{1}$ & upper layer depth & $250$m  \\
    $H_{2}$ & upper layer depth & $750$m  \\
    $\Delta \rho$ & density difference & $\rho_2 - \rho_1 =  1$ kgm$^{-3}$  \\
    $U$ & characteristic  velocity & $\tau_{0}/(\rho_0 H_{1} \beta_{0} L_y)$ = 0.02 m s$^{-1}$ \\
    $\tau_{0}$ & characteristic zonal wind stress & 0.1 Pa \\
    $\beta_{0}$ & planetary vorticity gradient & $2 \times 10^{-11}$ (ms)$^{-1}$\\
    $f_0$  & reference Coriolis parameter & $10^{-4}$ s$^{-1}$ \\
    $\epsilon_0$  & bottom friction coefficient & $10^{-7}$ s$^{-1}$ \\
    $g$ & gravitational acceleration & $9.8$ ms$^{-2}$ \\
    $g'$ & reduced gravity & $g \Delta \rho/\rho_0 = 0.01$ ms$^{-2}$ \\
    $\rho_0$ & reference density & $10^{3}$ kgm$^{-3}$ \\ \hline \hline
    $F_1$ & upper layer Froude number & $f_0^2L_y^2/( g' H_1 ) = 4000$ \\
    $F_2$ & lower layer Froude number & $f_0^2L_y^2/( g' H_2 ) = 4000/3$ \\
    $\beta$ & planetary vorticity factor & $\beta_{0} L_y^{2} / U = 1000$ \\
    $\tau$ & wind-stress parameter & $1.0$ \\
    $r_2$ & lower layer linear friction coefficient & $r_{2} = \epsilon_0 L_y/U = 5.0$ \\
    \hline
  \end{tabular}
  \caption{Set of model parameters used in the numerical study of the problem.}
  \label{tab:defp}
\end{table}

\subsection*{Acknowledgments:}
The authors are grateful for two anonymous reviewers for their insightful comments.

This work has been partially supported by the European Research Council (ERC) under the European Union’s Horizon 2020 research and innovation program (grant agreement No. 810370). This study was also supported by a Ben May Center grant for theoretical and/or computational research and by the Office of Naval Research Multidisciplinary University Research Initiative Grant N00014-20- 1-2023.

%
\bibliographystyle{plain}

\end{document}